\documentclass[a4paper, 11 pt]{article}
\usepackage{jheppub}
\usepackage{lineno}

\usepackage[utf8]{inputenc}
\usepackage[english]{babel}
\usepackage{subcaption}
\usepackage{float}
\usepackage{cleveref}

\title{\boldmath Allowable complex saddles with real charges in the gravitational index}

\author{Fabio Billiato}
\author{and Gianmarco Esposto}
\affiliation{INFN, Sezione di Padova, Via Marzolo 8, 35131 Padova, Italy}
\affiliation{Dipartimento di Fisica e Astronomia “Galileo Galilei", Università di Padova, \\
Via Marzolo 8, 35131 Padova, Italy}
\emailAdd{fabio.billiato@phd.unipd.it}
\emailAdd{gianmarco.esposto@phd.unipd.it}

\abstract{We present a new supersymmetric, non-extremal complexification of asymptotically $\mathrm{AdS}$ black holes, characterized by real charges and multiple complex horizons. We investigate whether these solutions define admissible supersymmetric saddles of the gravitational path integral by applying the Kontsevich-Segal-Witten (KSW) criterion, and compare the resulting gravitational constraints with the convergence conditions of the dual superconformal index. For most of the cases studied, the KSW criterion in the asymptotic region agrees with the index convergence, while the full bulk condition is more stringent and excludes saddles whose chemical potentials lie within the index-convergence domain.}

\renewcommand{\arg}{\mathrm{Arg\,}}
\def\be{\begin{equation}}
\def\ee{\end{equation}}
\def\rme{{\rm e}}
\newcommand{\diff}{\mathrm{d}}
\newcommand{\re}{\textrm{Re}}
\newcommand{\im}{\textrm{Im}}

\begin{document}
\maketitle
\flushbottom

\section{Introduction}

The Gravitational Path Integral (GPI), developed in the seminal work of Gibbons and Hawking~\cite{Gibbons:1976ue}, provides one of the main frameworks for studying quantum gravity, with applications ranging from quantum cosmology to black hole physics. In the latter context, the GPI has played a central role not only in establishing the thermodynamic properties of black holes and, more generally, of gravitational objects, but also in investigating subleading effects, including higher-derivative and quantum corrections.

In the semiclassical approximation, the path integral is expressed as a sum over gravitational saddles compatible with the boundary conditions. Of particular interest is the case in which supersymmetric boundary conditions are imposed, yielding the gravitational index. For asymptotically AdS spacetimes, the AdS/CFT correspondence identifies the gravitational index with a supersymmetric partition function, typically a superconformal index (SCI) \cite{Kinney:2005ej,Romelsberger:2005eg}, of the dual field theory. This correspondence has led to remarkable progress in the microscopic understanding of supersymmetric black holes. For instance, the large-$N$ behavior of the four-dimensional SCI has been shown to reproduce the AdS$_5$ black hole on-shell action, while an appropriate Legendre transform yields the Bekenstein-Hawking entropy \cite{Cabo-Bizet:2018ehj,Choi:2018hmj,Benini:2018ywd}. The agreement has subsequently been extended to four-derivative corrections \cite{Cassani:2022lrk,Cassani:2024tvk,Bobev:2022bjm} and to logarithmic corrections to the black hole action and entropy \cite{Cassani:2026cqp,David:2021qaa}.

The Euclidean saddles relevant for this comparison, in general, do not describe real black hole geometries. Instead, one considers smooth supersymmetric configurations with a compact Euclidean time circle. Such solutions are non-extremal and complex, and a real solution is recovered only in the extremal limit. Thus, reproducing the field-theory index requires extending the gravitational integration contour to a complexified space of metrics and matter fields.

In previous analyses \cite{Cabo-Bizet:2018ehj,Cassani:2019mms}, the relevant complex saddles were constructed by complexifying some of the parameters of the solution, while keeping the spacetime coordinates real. This choice results in solutions characterized by complex conserved charges. Although this is the most commonly used prescription, it is not the only possible choice compatible with supersymmetry. In this paper, we introduce a different complexification that keeps the conserved charges real while allowing the radial coordinate to take complex values. Once the reality condition on the radial coordinate is relaxed, different complex horizons emerge and can be treated on an equal footing. 

Black holes may in general possess more than one horizon, the simplest example being the outer and inner horizons of rotating or electrically charged solutions. Inner horizons share several formal properties with outer event horizons. In particular, they arise as roots of the blackening function $\Delta_r(r)$, which is related to the inverse radial component of the metric through $\Delta_r(r)\sim g_{rr}^{-1}$. This observation makes it possible to associate formal thermodynamic quantities, such as temperature, chemical potentials, and entropy, with each root. These quantities satisfy relations analogous to the standard first law \cite{Cvetic:1997uw,Cvetic:1997xv,Wu:2004yk,Cvetic:2018dqf,Detournay:2012ug,Castro:2012av}, although their physical interpretation is often subtle. For example, the conventional temperature associated with a Cauchy horizon is typically negative, and different interpretations of this feature have been proposed in the literature \cite{CurirSpin,Cvetic:1997uw,Cvetic:1997xv,Larsen:1997ge} (see also \cite{Strominger:1997eq,Guica:2008mu,Castro:2010fd,Nian:2020qsk,Hristov:2023sxg}).

The horizon structure is richer for asymptotically AdS black holes. In this case, the blackening function is generally a polynomial of higher degree in the radial coordinate than for the corresponding asymptotically flat solutions. It therefore admits additional roots $r_i$, some of which remain complex for every value of the physical parameters. We distinguish two classes of roots. We call \emph{outer horizons} those that can become real and positive in some limit (usually the extremal limit) and \emph{virtual horizons} those that always remain complex.
For virtual horizons, thermodynamic quantities are generically complex-valued but they nevertheless obey standard thermal relations \cite{Castro:2010fd,UniversalAreaProduct,Hristov:2023cuo}. 

Since in our supersymmetric complexification all horizons are complex away from extremality, it is natural to treat them on an equal footing. Importantly, each horizon may be chosen as the endpoint of a radial contour in the complexified spacetime. Choosing a given horizon therefore specifies the radial integration cycle on which the on-shell action is evaluated and, consequently, defines a distinct supersymmetric saddle in the semiclassical expansion of the GPI. As already emphasized, complex saddles can contribute to the GPI and this has proved essential for reproducing field-theory results. Nevertheless, it is important to determine which of them should be included in the gravitational path integral and which not. We investigate which complex horizons, and hence which associated saddles, should be regarded as physically allowable and which should instead be interpreted as mathematical artifacts.

In order to address this problem, we employ the Kontsevich-Segal-Witten (KSW) allowability criterion formulated in~\cite{Witten:2021nzp}, building on previous work~\cite{Kontsevich:2021dmb,Louko:1995jw}. According to the KSW criterion, a gravitational background is allowed in the GPI if it supports a well-defined quantum field theory. More specifically, the real parts of the kinetic terms must be positive, ensuring the convergence of the corresponding Gaussian functional integrals.

The KSW criterion has found important applications in quantum cosmology, where complex saddles play a prominent role. They arise, for instance, in semiclassical proposals for the wave function of the universe, an example being the no-boundary proposal~\cite{Hartle:1983ai,Hartle:2008ng,Lehners:2023yrj}, as well as in Lorentzian formulations based on Picard-Lefschetz theory, in which the original real integration contour is deformed into the complexified configuration space~\cite{Feldbrugge:2017kzv}. Applications of the KSW criterion in this context can be found in~\cite{Janssen:2024vjn,Hertog:2023vot,Hertog:2024nbh,Jonas:2022uqb}. 

More recently, the criterion has also been applied to complex black hole geometries~\cite{Chen:2023mbc,Chen:2022hbi,Ailiga:2025osa,Caminiti:2026efx}, and in particular to four- and five-dimensional complex supersymmetric black holes obtained through the standard complexification procedure~\cite{Chryssanthacopoulos,BenettiGenolini:2025jwe,BenettiGenolini:2026raa,Krishna:2026rma,Jones:2025gno}. In these examples, the KSW criterion was found to impose the same constraints on the chemical potentials as those required for convergence of the index. Moreover, the strongest restrictions arose in the asymptotic region, while the bulk analysis introduced no additional constraints.

The main goal of this paper is to apply the KSW criterion to our new supersymmetric non-extremal black holes, and to compare the resulting restrictions with the convergence conditions of the index. We find that, in both four and five dimensions, the KSW criterion excludes all virtual horizons, in agreement with the field-theory convergence conditions. For outer horizons,  the KSW criterion is more restrictive and rules out several saddles for which the index converges. Moreover, in contrast with the previously studied examples \cite{BenettiGenolini:2025jwe,BenettiGenolini:2026raa,Krishna:2026rma,Jones:2025gno}, the KSW criterion becomes more stringent in the bulk, excluding complex saddles that satisfy the asymptotic KSW condition. In cases where multiple unequal angular momenta are turned on, hence only in five dimensions, we also find that the KSW criterion applied to the asymptotic region already induces stronger constraints as compared to the index-convergence requirement. We interpret this result, along the lines of \cite{BenettiGenolini:2026raa}, as an indication that the formulation of the KSW criterion adopted in this work for the five-dimensional supergravity theory is too restrictive in a supersymmetric setup involving cancellations between bosonic and fermionic states.

The rest of the paper is organized as follows. In section~\ref{sec2}, we briefly review the origin of the convergence conditions for the SCI and the KSW criterion. In section~\ref{sec:charged-rotating-ads4}, we introduce the new complexification for the Euclidean supersymmetric non-extremal charged and rotating black hole in $\mathrm{AdS}_4$. We then apply the KSW criterion and compare the resulting constraints with those arising from convergence of the index. In section~\ref{sec:5d}, we carry out the analogous analysis in five dimensions. Finally, in section~\ref{sec:discussion}, we summarise and discuss our results.

\section{Complex supersymmetric saddles and index convergence}\label{sec2} 
In this section, we review the main ideas underlying the appearance of complex supersymmetric black hole solutions in the Euclidean gravitational path integral. Through holography, the contribution of these saddles can be compared with protected quantities in the dual field theory, such as the superconformal index. The convergence of the microscopic trace defining the index restricts the allowed values of the chemical potentials. On the gravity side, the same chemical potentials determine the boundary conditions of generally complex black hole geometries, whose allowability can be tested using the KSW criterion. Comparing these two sets of conditions will provide an important consistency check in the following sections.  
\subsection{The gravitational index}
Consider the Euclidean gravitational path integral with fixed asymptotically AdS boundary conditions, collectively denoted by $\mathcal{B}$:
\begin{align}
    \mathcal{Z}_{\mathrm{grav}}[\mathcal{B}]
    =
    \int_{\partial\mathcal{M}=\mathcal{B}}
    \mathcal{D} g\,\mathcal{D}\Phi\,
    e^{-I_{\mathrm{E}}[g,\Phi]}\,,
\end{align}
where $g$ is the metric, $\Phi$ collectively denotes the remaining fields of the theory, and $I_{\mathrm{E}}$ is the Euclidean action. Although this expression is formal, and no general non-perturbative definition of the gravitational path integral is presently known, it admits a semiclassical expansion around solutions of the Euclidean equations of motion satisfying the prescribed boundary conditions:
\begin{align}
    \mathcal{Z}_{\mathrm{grav}}[\mathcal{B}]
    \simeq
    \sum_{\alpha}
    e^{-I_{\mathrm{E}}[g_{\alpha},\Phi_{\alpha}]}\,.
\end{align}
The configurations $(g_{\alpha},\Phi_{\alpha})$ are candidate saddle points of the path integral. In general, they do not need to be real: complex solutions may contribute after an appropriate deformation of the original integration contour in the complexified space of fields.

In the context of the AdS/CFT correspondence, the gravitational partition function with boundary data $\mathcal{B}$ is identified with the partition function of the dual quantum field theory coupled to the corresponding background sources:
\begin{align}
    \mathcal{Z}_{\mathrm{grav}}[\mathcal{B}]
    =
    \mathcal{Z}_{\mathrm{QFT}}[\mathcal{B}]\,.
\end{align}
When the boundary conditions preserve supersymmetry, the field-theory partition function can be interpreted as a supersymmetric index. 

Consider, for example, asymptotically $\mathrm{AdS}_5$ Euclidean black hole solutions of five-dimensional minimal gauged supergravity. This theory admits consistent embeddings into type IIB string theory, for example through compactifications on suitable Sasaki-Einstein five-manifolds \cite{Gauntlett:2007ma,Buchel:2006gb}. These solutions are dual to states in the universal sector of four-dimensional $\mathcal{N}=1$ SCFTs defined on the boundary $\partial\mathcal{M}=S^1\times S^3$. The two angular momenta $J_1$ and $J_2$ are mapped to the Cartan generators of rotations on $S^3$, while the electric charge $Q$ of the black hole is mapped, up to a normalization, to the superconformal $U(1)_R$ charge $R$.

The corresponding chemical potentials are fixed by the asymptotic boundary data of the gravitational solution. In particular, the angular velocities $\Omega_1$ and $\Omega_2$, measured with respect to a non-rotating frame at infinity, are encoded in the off-diagonal components of the boundary metric. Schematically, the latter takes the form
\begin{align}\label{5dGeneralBoundaryGeom}
    \diff s_{\mathrm{bdry}}^2
    =
    \beta^2\diff \tau^2
    +
    \diff\theta^2
    +
    \sin^2\theta
    \left(
    \diff\phi-i\beta\Omega_1\diff \tau
    \right)^2
    +
    \cos^2\theta
    \left(
    \diff\psi-i\beta\Omega_2\diff \tau
    \right)^2.
\end{align}
Similarly, the electric potential $\Phi$ is encoded in the holonomy of the boundary $U(1)_R$ gauge field along $S^1$. The grand-canonical partition function of the dual field theory is therefore
\begin{align}
\label{trace1}
    \mathcal{Z}
    =
    \operatorname{Tr}_{\mathcal{H}_{S^3}}
    \left[
    \rme^{-\beta E+\beta\Omega_1J_1+\beta\Omega_2J_2+\beta\Phi R}
    \right].
\end{align}
This trace sums over the states of the SCFT on $S^3$, weighted
by their energy, angular momenta, and R-charge. Through the
state-operator correspondence, these states are associated with
local operators and organize into conformal families consisting
of a primary and its descendants. Conformal descendants are
generated by the translation operators $P_\mu$. We choose a supercharge $\mathcal{Q}$ such that
\begin{align}
\label{eq:BPSoperator}
    \{\mathcal{Q},\mathcal{Q}^{\dagger}\}
    =
    E-J_1-J_2-\frac{3}{2}R.
\end{align}
The BPS states selected by this supercharge thus satisfy $ E-J_1-J_2-\frac{3}{2}R=0$. Introducing the reduced chemical potentials
\begin{align}
    \omega_i=\beta(\Omega_i-1),
    \qquad
    \varphi=\beta\left(\Phi-\frac{3}{2}\right),
\end{align}
we can rewrite~\eqref{trace1} as
\begin{align}
\label{eq:reducedtrace}
    \mathcal{Z}
    =
    \operatorname{Tr}_{\mathcal{H}_{S^3}}
    \left[
    \rme^{-\beta\{\mathcal{Q},\mathcal{Q}^{\dagger}\}
    +\omega_1J_1+\omega_2J_2+\varphi R}
    \right].
\end{align}
Imposing the supersymmetry constraint \cite{Cabo-Bizet:2018ehj,Cabo-Bizet:2019osg}
\begin{align}
\label{susyconst2}
    \omega_1+\omega_2-2\varphi
    =
    2\pi i(1+2n_0),
\end{align}
it is possible to eliminate $\varphi$ and obtain the superconformal index,
\begin{align}
\label{traceSCI}
    \mathcal{I}_{n_0}(\omega_1,\omega_2)
    =
    \operatorname{Tr}_{\mathcal{H}_{S^3}}
    \left[
    \rme^{-\pi i(1+2n_0)R}
    \rme^{-\beta\{\mathcal{Q},\mathcal{Q}^{\dagger}\}}
    \rme^{\omega_1\left(J_1+\frac{R}{2}\right)
    +\omega_2\left(J_2+\frac{R}{2}\right)}
    \right].
\end{align}
The combinations $J_i+R/2$ commute with the chosen supercharge,
whereas the phase $\rme^{-\pi i(1+2n_0)R}$ changes sign between
states paired by $\mathcal{Q}$. Consequently, states with
$\{\mathcal{Q},\mathcal{Q}^{\dagger}\}>0$ cancel in the trace. The resulting protected trace receives
contributions only from BPS states and is independent
of $\beta$ at fixed $\omega_1$, $\omega_2$, and $n_0$. In order for the index to be convergent we must require the following conditions:
\begin{align}\label{indexconvergencecond}
\re \, \beta >0, \qquad \re \, \omega_1<0,\qquad \re\, \omega_2<0\,.    
\end{align}
The first condition ensures that the cancellation between bosons and fermions is well defined.  
The other two conditions can be understood by considering the conformal descendants generated by translations. We denote by $P_i^\pm$ the components of the translation generators whose action on a state shifts its quantum numbers according to
\begin{align}
\nonumber    P_1^\pm &: \quad
    (\delta E,\delta J_1,\delta J_2,\delta R)
    =(1,\pm1,0,0),\\
    P_2^\pm &: \quad
    (\delta E,\delta J_1,\delta J_2,\delta R)
    =(1,0,\pm1,0).
\end{align}
The operators $P_1^+$ and $P_2^+$ preserve  $\{\mathcal{Q},\mathcal{Q}^{\dagger}\}$. 
Starting from a BPS state with charges
$(E_0,J_{1,0},J_{2,0},R_0)$, their repeated action produces BPS descendants with charges
\begin{align}
    (E,J_1,J_2,R)
    =
    (E_0+n+m,J_{1,0}+n,J_{2,0}+m,R_0).
\end{align}
The contribution of a primary operator and its tower of descendants to the index is proportional to
\begin{equation}
\sum_{n,m\geq0}\rme^{n\omega_1+m\omega_2}\,,
\end{equation}
whose convergence requires the last two conditions in~\eqref{indexconvergencecond}.

\subsection{KSW criterion}
In the following we briefly review the original formulation of the allowability criterion introduced in \cite{Witten:2021nzp}, before presenting the alternative formulation developed in~\cite{BenettiGenolini:2026raa}, which will prove useful in the subsequent analysis.

Let $\mathcal{M}$ be a smooth manifold of dimension $d$, and let $g$ be a non-degenerate complex-valued symmetric metric on $\mathcal{M}$. The metric is said to be allowable if, at every point $x\in\mathcal{M}$ and for every non-zero real $p$-form $F_p$, the following inequality holds:
\begin{align}
\label{ksw1}
\re\left(
\sqrt{\det g}\,
g^{i_1j_1}g^{i_2j_2}\cdots g^{i_pj_p}
F_{i_1i_2\cdots i_p}
F_{j_1j_2\cdots j_p}
\right)>0\,,
\qquad
0\leq p\leq d\,.
\end{align}
It has been shown in \cite{Kontsevich:2021dmb} that this family of inequalities is equivalent to the existence, at each point $x\in\mathcal{M}$, of a basis of the real tangent space $T_x\mathcal{M}$ in which the metric takes the diagonal form
\begin{align}
g=\sum_{i=1}^{d}\lambda_i\,\diff y_i^2\,,
\end{align}
with non-zero complex coefficients $\lambda_i$ satisfying
\begin{align}
\label{ksw2}
\sum_{i=1}^{d}\left|\arg\lambda_i\right|<\pi\,,
\end{align}
where $\arg z\in(-\pi,\pi]$ denotes the principal value of the argument. 

It is worth emphasizing that the $\lambda_i$ appearing in~\eqref{ksw2} are not, in general, the eigenvalues of the matrix representing $g$. Since $g$ is a bilinear form, a change of real basis acts by congruence rather than by similarity. Thus, the diagonal form is obtained by finding a real invertible matrix $P$ such that
\begin{align}
g'=P^{T}gP
\end{align}
is diagonal, rather than by performing a transformation of the form $P^{-1}gP$.

A computationally convenient reformulation of the KSW criterion was presented in \cite{BenettiGenolini:2026raa}. It is particularly convenient for numerical applications, since it allows the criterion to be tested directly in an arbitrary real basis, without explicitly constructing the congruence transformation that diagonalizes the metric. At each point of $\mathcal{M}$, define
\begin{align}
A\equiv\re\left(\sqrt{\det g}\,g^{-1}\right).
\end{align}
Then $g$ satisfies the KSW criterion if and only if the following three conditions hold:
\begin{enumerate}
    \item $\det g\notin\mathbb{R}_{<0}$, so that the branch of $\sqrt{\det g}$ can be chosen with $\re \,\sqrt{\det g}>0$;

    \item $A$ is positive definite;

    \item the eigenvalues $\mu_i(gA)$ of the matrix $gA$ satisfy $\sum_{i=1}^{d}\left|\arg\mu_i(gA)\right|<\pi$.
\end{enumerate}
Depending on the problem under consideration, we will use either the original definition~\eqref{ksw1}, the equivalent diagonal characterization~\eqref{ksw2}, or the formulation in terms of the matrix $A$ and the eigenvalues of $gA$.

We end this section with an observation.
The KSW criterion should be imposed point-wise over the whole bulk geometry. In particular, in the asymptotic region the metric can be written in the Fefferman-Graham gauge as 
\begin{equation}\label{GeneralFGExpansion}
    ds^2 \sim \frac{dz^2+ds^2_{\rm bdry}}{z^2}\,, 
\end{equation}
where $z=0$ corresponds to the boundary and  $ds^2_{\rm bdry}$ is the dual CFT's background~\eqref{5dGeneralBoundaryGeom}. The KSW criterion applied to~\eqref{GeneralFGExpansion} only involves the induced boundary metric $ds^2_{\rm bdry}$ which, at finite temperature and angular velocity, it explicitly depends on $\beta,\Omega_i$ and is thus complex-valued assuming $\beta,\Omega_i \in \mathbb C$. As discussed in \cite{BenettiGenolini:2026raa}, building on previous considerations by \cite{Witten:2021nzp}, the KSW criterion applied to $ds^2_{\rm bdry}$ is equivalent to the requirement that the field theory partition function $\mathcal Z$ in \eqref{trace1} is well defined for general complex values of $\beta$ and $\Omega_i$.
 The corresponding convergence conditions can be obtained by generalizing the argument of \cite{Hawking:1999dp} and were worked out in \cite{BenettiGenolini:2026raa}. They read
\begin{equation}
    \re\,\beta >0\,, \qquad \left|\frac{\re\,(\beta\,\Omega_i)}{\re\,\beta}\right| <1\,,
     \qquad \longleftrightarrow \qquad 2\re \beta >-\re\, \omega_i >0\,. 
     \label{eq:nonBPSdescendantconvergence}
\end{equation}
The constraints in~\eqref{eq:nonBPSdescendantconvergence} are more stringent than the microscopic constraints arising from convergence of the index~\eqref{indexconvergencecond}. This stronger requirement is perhaps natural to expect, since the KSW criterion, in the formulation of~\cite{Kontsevich:2021dmb,Witten:2021nzp}, is not derived within a supersymmetric setup and may therefore fail to reproduce the constraints arising in the dual supersymmetric description, as indeed occurs in some of the examples discussed below.

\section{Charged, rotating AdS\texorpdfstring{$_4$}{} black hole}
\label{sec:charged-rotating-ads4}
In this section we study the charged rotating AdS$_4$ saddle of minimal gauged $\mathcal N=2$ supergravity and its complexifications. We first review the solution and its thermodynamic properties. We then distinguish two inequivalent complexifications of the supersymmetric family: one keeps the radial endpoint real by complexifying the mass and the charge of the solution, whereas the other keeps the mass and the conserved charges real, while allowing the roots of $\Delta_r$ to move into the complex plane. The latter parametrization will be used to compare the convergence domain of the boundary superconformal index with the KSW constraints on the bulk metric.

\subsection{General charged, rotating solution}

The bosonic sector of four-dimensional minimal gauged $\mathcal{N}=2$ supergravity is described by Einstein-Maxwell theory with a negative cosmological constant,
\begin{equation}
S \,=\, \frac{1}{16\pi G}\int \diff^4x \, \sqrt{-g}\left(R+\frac{6}{\ell^2}- F_{\mu\nu}F^{\mu\nu}\right),
\label{eq:4DMinimalGaugedAction}
\end{equation}
where $\ell$ is the AdS$_4$ radius.
The general electrically charged, rotating, asymptotically AdS$_4$ solution was originally found in~\cite{Carter:1968ks}, and its thermodynamics was analyzed in detail in~\cite{TDofKNADS_BH}. 
Adopting the parametrization of~\cite{Bobev:2019zmz}, the solution depends on three constants $(m,a,\delta)$ and can be written in Boyer-Lindquist-type coordinates\footnote{Relative to the conventions of~\cite{TDofKNADS_BH}, the parameters are related by
\[
m_{\rm there}=m_{\rm here}\cosh(2\delta)\,,\qquad
a_{\rm there}=a_{\rm here}\,,\qquad
q_{\rm there}=m_{\rm here}\sinh(2\delta)\,,\qquad
r_{\rm there}=r_{\rm here}\,.
\]
The radial coordinate used here is the shifted coordinate denoted by $\tilde r$ in~\cite{Bobev:2019zmz}; in terms of the unshifted coordinate $r$ of that reference, $\tilde r=r+2m\sinh^2\delta$.
}
$(t,r,\theta,\phi')$.
 
For later convenience, we work in a frame that is co-rotating with the horizon, achieved by introducing an angular coordinate $\phi=\phi'-\Omega_{\rm hor}t$, where $\Omega_{\rm hor}$ is the angular velocity at the horizon in the coordinate system $(\phi',t)$. After Euclidean continuation $t=-i\beta\tau$, and setting $\ell=1$ from now on, the metric and gauge field take the form: 
\begin{align} \label{bh4d}
\nonumber \diff s^2 ={}& W\left( \frac{\diff r^2}{\Delta_r} + \frac{\diff\theta^2}{\Delta_\theta} \right) + \frac{\Delta_r \Delta_\theta}{B \Xi^2}\, \beta^2\, \diff\tau^2 \nonumber\\
\nonumber &+ \sin^2\theta\, B \left( \diff\phi - i a\, \frac{ \Delta_r\left(r_+^2+a^2\cos^2\theta\right) + \Delta_\theta\left(r^2+a^2\right)\left(r^2-r_+^2\right) }{ \left(r_+^2+a^2\right) B W \Xi }\, \beta\, \diff\tau \right)^2,\\
\mathcal{A} ={}& \frac{m r \sinh 2\delta }{W\Xi} \left[-i\beta \left(\Delta_\theta\, \diff\tau - a\sin^2\theta\,\Omega\, \diff\tau\right) - a\sin^2\theta\, \diff \phi\right] + i\beta \Phi\, \diff \tau \,,
\end{align}
where we added a constant term $i\beta\Phi\, \diff \tau$ to the gauge field to ensure its regularity at the horizon. The metric functions read
\begin{align}
\Delta_r &= (r^2+a^2)(1+r^2) -2mr\cosh2 \delta +m^2\sinh^2 2\delta\,, \nonumber\\[4pt]
\Delta_\theta &= 1-a^2\cos^2\theta\,, \qquad
W = r^2+a^2\cos^2\theta, \nonumber\\[4pt]
\Xi &= 1-a^2\,, \qquad
B = \frac{ \Delta_\theta (r^2+a^2)^2 - a^2\sin^2\theta\,\Delta_r }{ W\Xi^2 }\,.
\end{align}
Denoting by $r_+$ the largest positive root of $\Delta_r$, the inverse temperature $\beta$, angular velocity $\Omega$, and electrostatic potential $\Phi$ of the black hole are:
\begin{equation}
\beta= \frac{4\pi (r_+^2+a^2)}{\Delta_r'(r_+)}\,, \qquad \Omega = \frac{a(1+r_+^2)}{r_+^2+a^2}\,, \qquad \Phi = \frac{m r_+ \sinh \,2\delta}{r_+^2+a^2}\,.
\label{eq_1Q1J4DCHemPot}
\end{equation}
In these co-rotating coordinates, the metric~\eqref{bh4d} is static at the horizon, with coordinates obeying untwisted periodicity conditions
\begin{align}
    \phi \sim \phi+2\pi, \qquad \tau \sim \tau+1.
\end{align}
The rotation parameter must satisfy $a^2<1$, so that both $\Xi$ and $\Delta_\theta$ remain positive and non-vanishing for all values of $\theta$.  
Without loss of generality, we can assume that all parameters are non-negative \cite{Bobev:2019zmz}.

The solution is characterized by three conserved charges: mass $M$, angular momentum $J$ and electric charge $Q$ \cite{Bobev:2019zmz}:
\begin{equation}
M =\frac{m \cosh\, 2\delta}{G\,\Xi^2}\,, \qquad J = \frac{am\cosh \, 2\delta}{G\,\Xi^2}\,,\qquad Q = \frac{m\sinh \, 2\delta}{G\,\Xi}\,,
\label{eq:1Q1J4DCharges}
\end{equation}
and the Bekenstein-Hawking entropy associated with the outer horizon is
\begin{equation}
S = \frac{\pi(r_+^2+a^2)}{G\,\Xi}\,.
\label{eq:1Q1J4DEntropy}
\end{equation}
These quantities satisfy the first law of thermodynamics 
\be\label{eq:first_law}
\diff M = T\,\diff S+\Omega \, \diff J+\Phi \, \diff Q\,.
\ee
Moreover, by evaluating the Euclidean on-shell action $I$, one can also verify that the quantum statistical relation holds
\be\label{eq:QSR}
I = -S + \beta\, M -\beta \,\Omega\,J-\beta\,\Phi\, Q\,,
\ee
which identifies $I/\beta$ with the Gibbs thermodynamic potential.

\paragraph{Complex horizons and integration contours.}
We now consider a class of complexifications where the event horizon of the black hole~\eqref{bh4d} is allowed to become complex-valued  $r_+\in \mathbb C$, while the $a,m,\delta$ parameters are kept real.  Let us clarify the nature of these horizons. 

Killing horizons are defined as null hypersurfaces on which the norm of a given Killing vector $\xi$ vanishes. 
For the black hole solution~\eqref{bh4d}, the Killing horizons coincide with the roots of $\Delta_r$, which is a quartic polynomial of $r$ with real coefficients.
For general values of the $a,m,\delta$ parameters, the condition $\Delta_r=0$ may not have real solutions, an example being given by super-extremal configurations with $m<m_{\rm ext}$ \cite{Caldarelli:1998hg}, in which the event horizon and Cauchy horizon roots, $r_+$ and $r_0$, become a pair of complex conjugate roots\footnote{In this sense one can understand sub-extremal black holes as having a pair of real roots $r_+>r_0>0$, at extremality $r_+=r_0$ and $\Delta_r$ has a double root. For super-extremal configurations $r_+$ and $r_0$ split again, but along the imaginary axis rather than along the real axis, as they become complex-valued.} $(r_+,\bar r_+)$ of $\Delta_r$. The supersymmetric but non-extremal configurations that we will discuss are also of this form. 

Importantly, the relevant thermodynamic quantities associated with Killing horizons, such as chemical potentials and Bekenstein-Hawking entropy, can also be defined for complex-valued horizons. This is simply achieved by analytically continuing their expressions to $r_+\in \mathbb C$. In order to do so, one also has to trade one parameter, usually $m$, for $r_+$ using the condition $\Delta_r(r_+)=0$. The first law of thermodynamics~\eqref{eq:first_law} continues to hold, and one can use~\eqref{eq:QSR} to define a complexified Gibbs potential, which again is just given by the analytic continuation of its corresponding real function of $r_+$. Note that the charges~\eqref{eq:1Q1J4DCharges} are always kept real. 

We can go further and interpret black hole solutions with complex horizons as genuine (complex) saddle points of the gravitational path integral. We do so by first introducing a complex manifold  $\mathcal M_{\mathbb C}$ obtained by complexifying the radial coordinate $r$ in the geometry~\eqref{bh4d}. Then we consider a real section $\mathcal M_{r_+} \subset \mathcal M_{\mathbb C}$ obtained by choosing a smooth curve $r(s)$ parametrized by a real parameter $s\in [\rho,\infty)$ such that:
\begin{equation}
    r(\rho) = r_+\,, \qquad \lim_{s\to+\infty} \text{Re} \,r(s) = +\infty \,, \qquad \lim_{s\to +\infty} \im \, r(s) = 0\,,
    \label{eq:ContourBoundaryConditions}
\end{equation}
meaning that the curve ends on the complex horizon $r_+$, and reaches the asymptotic region $r\to+\infty$ along the real axis, where the metric retains its usual AdS$_4$ form. This requirement allows us to provide an holographic interpretation for the given saddle point.  
With this choice, the metric~\eqref{bh4d} gives an asymptotically AdS$_4$ complex metric defined on the real manifold $\mathcal M_{r_+}$, parametrized by real coordinates $(s,\tau,\theta,\varphi)$. 
$\mathcal M_{r_+}$ has the usual cigar geometry associated with Euclidean black holes. In order to see this, we expand the metric~\eqref{bh4d} close to $r(\rho)=r_+$ with
\begin{equation}
    s-\rho= R^2 \ll 1\,,
\end{equation}
so that
\begin{equation}\label{eq:nh3}
    \diff s^2 \sim\frac{4  W(r_+) \, r'(s)|_{s=\rho}}{\Delta_r '(r_+)} \left(\diff R^2+4 \pi^2 R^2 \diff \tau^2\right)
    +\frac{W(r_+)}{\Delta_\theta }\diff \theta^2
    +\frac{ \sin ^2\theta \,\Delta_{\theta} \left(a^2+r_+^2\right)^2}{\Xi^2 W(r_+)} \diff \phi^2\,,
\end{equation}
in which the $R,\tau$ coordinates describe the usual regular $\mathbb R^2$ factor. Note that this allows us to interpret the complexified chemical potentials, associated with $r_+$, as arising from requiring regularity of the near-horizon metric as usual.\footnote{Indeed, remember that $t = -i\beta\tau$ and $\tau\sim \tau+1$. Had we used a different continuation we would have obtained a conical singularity at $r=r_+$.} In order to obtain a parametrization of $\mathcal M_{r_+}$ in terms of real coordinates, we have to consider the co-rotating frame in which $\tau\sim\tau+1$ and $\phi\sim\phi+2\pi$. Had we used static-at-infinity coordinates $(\tau,\phi')$ we would have been forced to introduce a complex angular coordinate, in order to ensure the twisted regularity condition $\tau\sim\tau+1$ and $\phi' \sim \phi' +i\Omega\beta$ at $r=r_+$. Finally, note that the near-horizon metric depends on the complex parameter
\begin{equation}
    \kappa = r'(s)|_{s=\rho} \in \mathbb C\,,
\end{equation}
which describes the direction in the complex $r$-plane along which the horizon $r_+$ is approached by the curve $r(s)$. The parameter $\kappa$ only appears as an overall factor in front of the $\mathbb R^2$ term, hence the same complex value of $\beta$ is associated with each path. 

These are the classes of complex metrics that we are going to consider. Similar complexifications have also been considered in \cite{Witten:2021nzp} as simple ways of constructing complex metrics, and in \cite{Chen:2023mbc,Chen:2022hbi}. Note that we could choose different curves satisfying the boundary conditions~\eqref{eq:ContourBoundaryConditions}. Following \cite{Witten:2021nzp}, we assume that different contours, possibly associated with different values of $\kappa$, define equivalent saddles whenever they can be smoothly deformed into one another, i.e. whenever they arise from homotopic immersions of $\mathcal{M}_{r+}$ into $\mathcal{M}_{\mathbb C}$.\footnote{One should also consider singular points in $\mathcal M_{\mathbb C}$. In particular, two curves are inequivalent if any continuous deformation connecting them necessarily crosses a singular point, leading to inequivalent saddles.}

We can compute the on-shell action associated with the solution~\eqref{bh4d} on $\mathcal M_{r_+}$ in the usual way \cite{Gibbons:1976ue} as $r=r_+$ still represents a bolt of $\xi$. For contours along which $r(s)$ does not pass through any singular point of $\mathcal M_\mathbb C$, and considering the analytic properties of $I$ as a function of $r_+$, the result one would get is precisely given by the analytic continuation of $I$ to $r_+\in \mathbb C$. This completes the discussion and allows us to interpret the on-shell action $I$ as related to the Gibbs potential of the solution with complex chemical potentials $\beta,\Omega,\Phi$, which upon imposing the supersymmetric constraints, should reproduce the superconformal index of the dual field theory. 

Before discussing supersymmetry, we elaborate further on the possible saddles that can be relevant in our setup.

\paragraph{Outer and virtual horizons.}
The previous discussion can be generalized to any root $r_i$ of the $\Delta_r(r)$ polynomial. Indeed, the considerations that have been made above only rely on the property that $\Delta_r(r_i)=0$. Hence, any root $r_i$ is a candidate endpoint, provided that the induced metric and gauge field are non-degenerate and regular on the corresponding integration cycle. To each root one can associate a regular complex metric on the real manifold $\mathcal M_{r_i}$ describing a candidate saddle point of the gravitational path integral with chemical potentials and on-shell action in which $r_+$ is substituted with the given root $r_i$.

As $r_i$ are roots of a quartic polynomial, their expressions in terms of the $a,m,\delta$ parameters are quite cumbersome. However, by rewriting $\Delta_r$ as
\begin{equation}
    \Delta_r(r) = \prod_{i=1}^4 (r-r_i)\,,
\end{equation}
one finds that the $r_i$ roots satisfy the following relations
\begin{equation}
\begin{cases}
\ \sum_{i=1}^4r_i = 0\,, \\[1mm]
\ \sum_{i<j} r_i\,r_j  = a^2+1\,, \\[1mm]
\ \sum_{i<j<k}r_i\,r_j\,r_k = 2m\,\cosh 2\delta\,, \\[1mm]
\ r_1\,r_2\,r_3\,r_4  = a^2 +m^2\sinh^2 2\delta\,.
\end{cases}
\label{eq:1Q1J4DRootsRelations}
\end{equation}
Assuming that all roots $r_i$ are real, the first equation in~\eqref{eq:1Q1J4DRootsRelations} implies that at least one root must be negative. 
However, $\Delta_r$ is strictly positive for $r<0$ and $m>0$, therefore there are no real negative zeroes of $\Delta_r$. It follows that there must always be at least one pair of complex conjugate roots ($r_-$,$\bar r_-$), with negative real part, which can never become real. We denote these roots as \emph{virtual horizons}.   
The other two solutions can either be both real (and positive), corresponding to the real event $r_+$ and Cauchy $0<r_0<r_+$ horizons, or organize in a complex conjugate pair $(r_+,\bar r_+)$ corresponding to the \emph{outer horizons}. At extremality $r_+=r_0 \in \mathbb R$.

As we will now show, in the supersymmetric but non-extremal regime all four horizons become complex. This naturally raises the question of whether the virtual horizons can themselves define supersymmetric saddle points and hence contribute to the index.

\subsection{Supersymmetry and complexification}\label{sec:AdS4_susy}
We now consider the supersymmetric, but non-extremal, limit of the solution~\eqref{bh4d} providing the complex gravitational solutions contributing to the gravitational index, dual to the superconformal index of the three-dimensional $\mathcal{N}=2$ SCFT on $S^1\times S^2$ \cite{Bobev:2019zmz,Choi:2019zpz,Nian:2019pxj}. We discuss our choice of complexification and how it differs from what has already been considered in the literature. 

 Supersymmetry for the solution~\eqref{bh4d} was first studied in~\cite{Caldarelli:1998hg,Kostelecky:1995ei}. It was found that the parameters must satisfy the relation
\be\label{eq:susycond_4d}
a \,=\, \frac{2}{\rme^{4\delta}-1}\,,
\ee
which we use to eliminate $\delta$.
Upon imposing~\eqref{eq:susycond_4d}, $\Delta_r$ reduces to a product of two second order polynomials $\Delta_r=\Delta_r^+\Delta_r^-$ where
\begin{equation}\label{factorized_Delta}
\Delta_r^{\pm}= r^2\mp ir \left(1+a\right) - a \pm\frac{i m}{a\left(1+\frac{2}{a}\right)^{1/2}}\,.
\end{equation}
Imposing $\Delta_r(r_i)=0$ allows us to trade the mass parameter $m$ for any given root $r_i$
\begin{align}
    m=\sqrt{a(2+a)}(a\pm i r_i)(r_i\mp i),
    \label{mass-root-relation}
\end{align}
where the $\pm$ sign is related to which of $\Delta_r^+$ or $\Delta_r^-$ vanishes when evaluated at $r=r_i$. Note that we are not making any assumptions on the reality of the parameters yet. In terms of $(a,r_i)$, the chemical potentials read
\begin{equation}
    \beta = \mp\frac{2 \pi i  \left(a^2+r_i^2\right)}{(a\pm 2  i \, r_i+1) \left(a-r_i^2\right)}\,,
    \qquad \Omega=\frac{a(1+r_i^2)}{a^2+r_i^2} \,,\qquad
    \Phi = \frac{r_i (r_i\mp i)}{a\mp i r_i}\,,
    \label{beta-phie-ar}
\end{equation}
while the associated field theory variables are given by
\begin{align}
    \omega &= \beta(\Omega-1)
    =\pm\frac{2 \pi i  (a-1)}{a\pm2 i \,r_i+1}\,,
    &
    \varphi &= \beta (\Phi-1)
    =\pm\frac{2 \pi i  (a\pm i r_i)}{a\pm2 i\, r_i+1}\,,
    \label{field-theory-potentials}
\end{align}
and satisfy the supersymmetric constraint 
\begin{align}
    \beta\bigl(1+\Omega-2\Phi\bigr) = \mp 2\pi i\,,\qquad\omega-2\varphi=\mp 2\pi i\,.
    \label{susy-chemical-constraint}
\end{align}
This discussion is valid for any root $r_i$ of the $\Delta_r$ polynomial. It follows that all four roots $r_i$ are associated with candidate supersymmetric saddle points of the GPI, potentially contributing to the index, and should be considered in our analysis.

These saddles are generically complex as can be seen by considering that~\eqref{mass-root-relation} has complex coefficients. More precisely, requiring that $\Delta_r$ has at least one real root $r_+ = r_+^\star>0$ while also keeping $a,m \in \mathbb R$, imposes the additional constraint
\begin{equation}
    m^\star = a(1+a)\sqrt{2+a}\,,
    \label{eq:BPSExtramlityConstr}
\end{equation}
which corresponds to
\begin{equation}
    r_+^\star = \sqrt a\,, \qquad r_-^\star = -\sqrt a+ i(a+1)\,,\qquad  \bar{r}_-^\star = -\sqrt a- i(a+1)
\end{equation}
where the real event horizon at $r=r_+^\star$ is a double root of $\Delta_r$. Thus, the corresponding solution has $\beta^{-1} \to 0$ and represents the real supersymmetric and extremal (BPS) configurations admitting a regular Lorentzian continuation. Conversely, imposing supersymmetry outside of the BPS locus requires complexifying at least one parameter among $a,m,r_i$. A generic non-extremal supersymmetric solution is associated with a complex Euclidean saddle of the gravitational path integral. If this saddle is associated with a complex horizon $r_i$ it is given by the solution on $\mathcal M_{r_i}$ as discussed above.

These saddles depend on two independent complex (or four real) field theory parameters $\beta,\omega$. Correspondingly, one should consider black hole solutions where the parameters $a,m,r_i$ are complex and constrained by~\eqref{mass-root-relation}, so as to obtain four independent parameters related to $\beta,\omega$ via~\eqref{beta-phie-ar} and~\eqref{field-theory-potentials}. Studying this four-parameter family of solutions in full generality is difficult due to the complicated expressions one should work with. In practice a common choice is to reduce to a lower-dimensional set by imposing additional reality conditions on the parameters. We already saw that requiring all parameters to be real produces the one-parameter family of BPS solutions. Thus, we can impose up to two constraints and consider a two-parameter family.

In previous analyses~\cite{Cabo-Bizet:2018ehj,Cassani:2019mms,Bobev:2019zmz} the choice was to keep the rotation parameter(s) and $r_i = r_+>0$ real, while complexifying $m$ (and other parameters in higher dimensions). This is the same choice adopted in recent studies on the allowability criteria for complex metrics \cite{Chryssanthacopoulos,BenettiGenolini:2025jwe,BenettiGenolini:2026raa,Krishna:2026rma} and allows us to keep a real radial coordinate $r$. However, other choices can be made. 

\paragraph{Supersymmetric solutions with real charges.}
Here we adopt a complementary prescription: we keep the parameters $a,m,\delta$ real, while all horizons $r_i$ are complex outside of the extremal solutions. With this choice, the conserved charges $M,J,Q$ remain real. The resulting complexification is therefore fully carried by the contour in the complex $r$ space. 

Additionally, with $a,m\in \mathbb R$  the two quadratic factors~\eqref{factorized_Delta} have complex conjugate coefficients. Thus, the four roots $r_i$ organize in complex conjugate pairs $(r_+,\bar r_+)$ and $(r_-,\bar r_-)$ as happens for super-extremal black holes. As $\Delta_r^\pm$ remain complex-valued also in the BPS limit, it follows that
\be
\Delta_r^+\,=\,(r -r_-)(r -r_+) \,, \qquad \Delta_r^- = \bar \Delta_r^+ = (r-\bar r_+)(r-\bar r_-)\,.
\ee
Rewriting $r_\pm = \rho_\pm + i\sigma_\pm$ and solving the equation $\Delta_r^+(r_\pm) = 0$ one finds
\begin{align}
\rho_{\pm}={}&\pm \frac{1}{2\sqrt2}\left[-(1-a)^2+\sqrt{(1-a)^4+\frac{16m^2}{a(a+2)}}\,\right]^{1/2}\,, \nonumber\\[2mm]
\sigma_{\pm}={}& \frac{(1+a)}{2}\mp\frac{1}{2\sqrt 2}\left[(1-a)^2+\sqrt{(1-a)^4+\frac{16m^2}{a(a+2)}}\,\right]^{1/2} \,.
\label{roots-rho-sigma}
\end{align}
The same expressions are found for the complex conjugated roots given by $\bar r_\pm = \rho_\pm -i\sigma_\pm$ upon imposing $\Delta_r^-(\bar r_\pm)=0$.
In the BPS limit, i.e. by imposing~\eqref{eq:BPSExtramlityConstr}, one can check that $r_+=\bar r_+ = \sqrt a$ and $r_- = r_-^\star$ as expected.

\begin{figure}[htbp]
\centering
\includegraphics[width=1\textwidth]{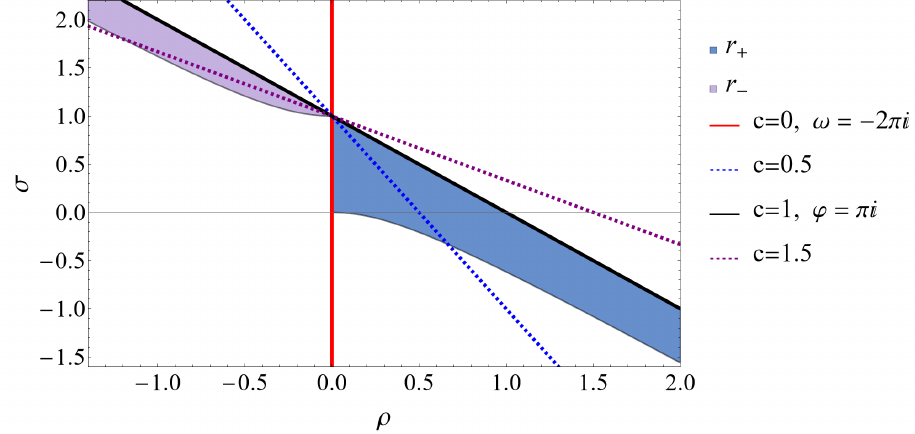}
\caption{Horizon domain in the $\rho,\sigma$ plane. The blue region corresponds to values associated with $r_+$ while the violet region to $r_-$. The dotted lines correspond to the loci where the supersymmetric chemical potentials $\omega,\varphi$ are constant, i.e. to the lines corresponding to $\frac{\rho}{1-\sigma} = c$.}
\label{fig:figure1}
\end{figure}

From~\eqref{roots-rho-sigma}, it is clear that
\begin{equation}
 \rho_+ \ge 0\,, \qquad \rho_- \le 0\,,
 \label{rhoRange}
\end{equation}
while the constraint $0 \leq a <1$ limits the range of the $\sigma_\pm$ parameter. To see this, it is useful to write $\sigma_\pm$ as a function of $a$ and $\rho_\pm$ as
\be
\sigma_\pm = \frac{1}{2}\left(1+a\mp\sqrt{(1-a)^2+4\rho_\pm^2}\right)\,,
\label{sigma-a-rho}
\ee
which are monotonically increasing functions of $a$ for all values of $\rho_\pm$. It follows that:
\be
0 \leq a < 1 \quad \longrightarrow \quad
    \frac{1}{2}\left(1\mp\sqrt{1+4\rho_\pm^2}\right) \leq \sigma_\pm < 1-\rho_\pm\,.
    \label{sigmaRange}
\ee
In the rest of this section, we refer to the region defined by this condition as the \emph{horizon domain}. 
  The $r_\pm$ domain in the $(\rho,\sigma)$ plane is shown in figure~\ref{fig:figure1}. Horizons with $\rho_{\pm}=0$ must be discarded, since from~\eqref{sigma-a-rho} they have $\sigma_+=a$ in which case $W$ vanishes at $\theta=0$ leading to a singular metric.

From now on we will use $\rho$ and $\sigma$ as parameters, suppressing the $\pm$ label and using the sign of $\rho$ to distinguish the two branches: negative values of $\rho$ describe virtual horizons, whereas positive values describe outer horizons. The $a,m$ parameters read
\begin{align}
    a&=\frac{-\rho^2-\sigma+\sigma^2}{-1+\sigma}, \nonumber\\
    m&=\frac{|\rho|\,\left(\rho^2+(-1+\sigma)^2\right)
    \sqrt{(\rho^2+\sigma-\sigma^2)(\rho^2-(-1+\sigma)(2+\sigma))}}{\left( -1+\sigma \right)^2}\,.
    \label{am-rhoplus-sigmaplus}
\end{align}
While the thermodynamic potentials are given by
\begin{align}
\beta={}&\frac{2 \pi i\,  \rho  \bigl(\rho ^3-\rho  \sigma ^2+\rho +2 i (\sigma -1)^2 \sigma \bigr)}{\sigma  \bigl(\rho ^2-2 i\, \rho  (\sigma -1)+(\sigma -1)^2\bigr) \bigl(-\rho ^2-2 i \rho  (\sigma -1)+\sigma ^2-1\bigr)}\,, \nonumber\\[1mm]
\omega={}& \frac{2 \pi i \bigl(\rho^2 -(\sigma-1)^2\bigr) }{\rho ^2-2 i \rho  (\sigma -1)+(\sigma -1)^2}\,,
\label{chempotrhosigma}\\[1mm]
\varphi={}& \frac{2 \pi i \, \rho  \bigl(\rho -i(\sigma -1)\bigr)}{\rho ^2-2 i \rho  (\sigma -1)+(\sigma -1)^2}\,. \nonumber
\end{align}
These are appropriate for both the outer and virtual horizons $r_\pm$, which are differentiated by the domains of $\rho$ and $\sigma$ given in eqs.~\eqref{rhoRange} and~\eqref{sigmaRange}. The corresponding expressions for the conjugated horizons $\bar r_\pm$, written in terms of $\rho$ and $\sigma$, are obtained by complex conjugation of the quantities appearing in~\eqref{chempotrhosigma}.

From~\eqref{chempotrhosigma} we see that the chemical potentials $\omega(\rho,\sigma)$ and $\varphi(\rho,\sigma)$ obey
\begin{align}
\omega(\rho,\sigma) &= \omega\left(\rho(1-\sigma)^{-1},0\right) = \omega^\star\left(\rho(1-\sigma)^{-1}\right)\,, \notag \\
\varphi(\rho,\sigma) &= \varphi\left(\rho(1-\sigma)^{-1},0\right) = \varphi^\star\left(\rho(1-\sigma)^{-1}\right)\,,
\label{eq:4dSusyChempotScaling}
\end{align}
where $\omega^\star(\rho)$ and $\varphi^\star(\rho)$ denote the restriction of the chemical potentials to the locus $\sigma=0$ and read
\be
\omega^\star = \frac{2\pi i(\rho^2-1)}{\rho^2+2i\rho+1}\,, \qquad \varphi^\star = \frac{2\pi \rho(i\rho-1)}{\rho^2+2i\rho+1}\,.
\label{eq:4dBPSChempotrhosigma}
\ee
For $0<\rho<1$, these are the expressions associated with the regular supersymmetric and extremal BPS black holes with $\rho=\sqrt a<1$. For $\rho >1$ these expressions are associated with non-regular BPS black holes with $a>1$.

The eqs.~\eqref{eq:4dSusyChempotScaling} show that $\omega$ and $\varphi$ depend only on the ratio $c=\frac{\rho}{1-\sigma}$. They are constant along lines with fixed $c$ (depicted as dotted lines in figure~\ref{fig:figure1}), with values given by $\omega^\star(c),\varphi^\star(c)$. Note that, at the singular point $\rho=0$ and $\sigma=1$, the chemical potentials do not acquire a definite value which instead depends on how the singular point is approached. This is evident from figure~\ref{fig:figure1}.

Moving along the constant-$c$ lines in the $\rho,\sigma$ space corresponds to departing from a given BPS solution by changing $\beta$ while keeping $\omega$ and $\varphi$ fixed to their corresponding BPS values $\omega^\star,\varphi^\star$. In general, our complexification is such that $\omega$ and $\varphi$ only take values on the BPS locus given by the one-dimensional curve, in the complex $\omega,\varphi$ space, defined by the parametrization given by~\eqref{eq:4dBPSChempotrhosigma}. This property can be actually understood as arising from the requirement of keeping real conserved charges and can be derived just from general considerations on the extremization procedure \cite{Cassani:2019mms,Bobev:2019zmz}, with no need to specify a given complexification of the black hole parameters. We explicitly prove this in Appendix \ref{app:RealChargExtrPrinc}. This implies that any such choice that leads to real charges must also imply that the chemical potentials remain on the BPS locus.

This differs from the complexification where one keeps $r_+$ real. In this case, moving away from a BPS solution by requiring $|\beta|<\infty$ always detunes $\omega$ and $\varphi$ away from their BPS locus. This explicitly shows that the two complexifications are inequivalent, as they explore different two-dimensional sections of the whole space of supersymmetric solutions. These sections only meet for $\beta^{-1}\to 0$, i.e. in the BPS limit, where both $r_+$ and the black hole parameters are real.

As a final comment we note that, as seen in figure~\ref{fig:figure1}, all the points associated with the outer horizon domain can be reached by lines with $0<c<1$ passing through a physical BPS solution, at $\sigma=0$ and  $0<\rho<1$. Instead, points associated with $r_-$ are reached by lines with $c>1$ which intersect the real $\sigma=0$ axis at unphysical configurations with $\rho>1$, corresponding to $a>1$. This property automatically excludes the virtual horizons from the set of allowed saddles.

\subsection{Allowability of the complex solution}\label{sec:3.3}
We now study the allowability criteria on the above complex solutions. In particular, we ask which of the four roots $(r_\pm,\bar r_\pm)$ can serve as the endpoint of an admissible contour in the complexified radial plane, thus defining an allowed supersymmetric saddle. We compare the convergence conditions of the superconformal index in the dual field theory with the KSW admissibility criterion imposed on the bulk metric.

\begin{figure}[htbp]
\centering
\includegraphics[width=0.56\textwidth]{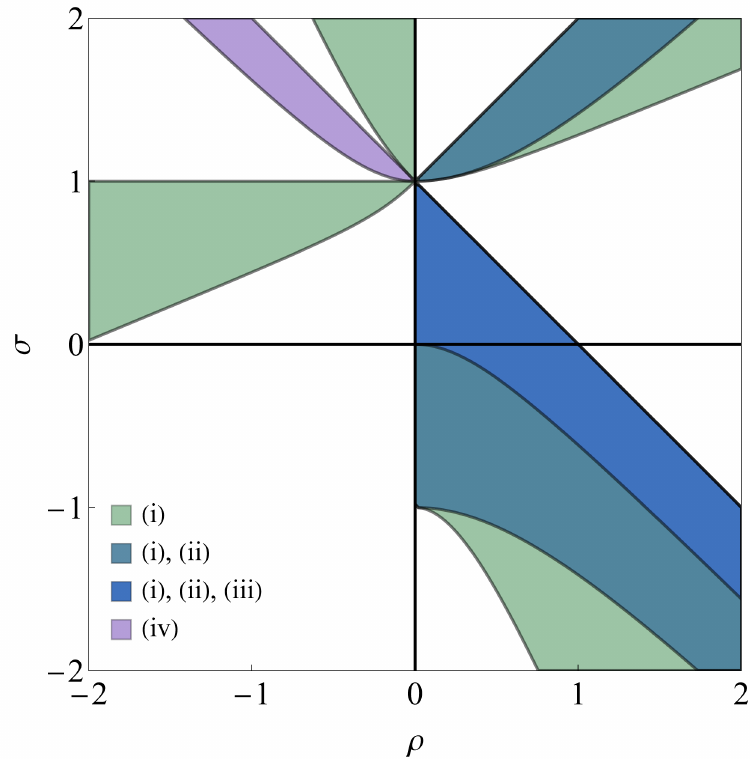}
\caption{Regions in the $(\rho,\sigma)$ plane selected by the field-theory and gravitational constraints. Points are classified according to whether they satisfy (i) the index-convergence conditions, (ii) the asymptotic KSW criterion, (iii) belong to the outer horizon domain, and (iv) belong to the virtual horizon domain. The different colored regions indicate the subsets of parameter space that satisfy the corresponding combinations of these constraints.
}
\label{fig:figure2}
\end{figure}

\paragraph{Boundary convergence.}\label{subsubsec:outer-scft-convergence}
Analogously to the five-dimensional case discussed in
section~\ref{sec2}, one can define a superconformal index for the three-dimensional SCFT dual to the asymptotically AdS$_4$ black hole \cite{Bobev:2019zmz}. The construction uses an $\mathcal{N}=2$ superconformal algebra and takes the form of a graded trace over the Hilbert space on $S^2$ \cite{Bhattacharya:2008zy,Imamura:2011su}.
The dual theory depends on the higher-dimensional embedding; specific examples can be found in \cite{Aharony:2008ug,Guarino:2015jca, Fluder:2015eoa}. We impose the following convergence conditions
\begin{align}
\label{rebetapos}
    \re\,\beta>0\,,
    \qquad
    \re\,\omega<0\,.
\end{align}
Expressing $\beta,\omega$ using~\eqref{chempotrhosigma}, the above conditions translate to:
\begin{align}
\nonumber \mathrm{Re}\,\beta=&  -\frac{4\pi\rho\,(\sigma-1)^2\Bigl(\rho^4-2\rho^2(\sigma-1)(3\sigma-2)+(\sigma-1)^2(\sigma^2-1)\Bigr)}{\Bigl(\rho^4+(\sigma-1)^2\bigl(6\rho^2+(\sigma-1)^2\bigr)\Bigr)\Bigl(\rho^4+2\rho^2(\sigma-1)(\sigma-3)+(\sigma^2-1)^2\Bigr)}>0\,,\\
\re\, \omega=& \,\frac{4 \pi  \rho  (\sigma -1) \left((\sigma -1)^2-\rho ^2\right)}{\rho ^4+6 \rho ^2 (\sigma -1)^2+(\sigma -1)^4}<0\,.
\label{rebetaexplicit}
\end{align}
In  figure~\ref{fig:figure2} the region of convergence of the index is shown in the $(\rho,\sigma)$ plane. For the outer horizons, the conditions above are always satisfied, and in particular, for $\rho>0$:
\begin{align}
    \textbf{outer horizon domain} \subset \textbf{index convergence}
\end{align}
For the virtual horizons $r_-$, the inequalities~\eqref{rebetaexplicit} are never satisfied as expected from the observation in eq.~\eqref{eq:4dSusyChempotScaling}. Therefore, convergence of the superconformal index rules out all saddles associated with the virtual horizons.

\paragraph{Asymptotic KSW analysis.}
\label{asranalysis}
We now compare the boundary convergence conditions with the KSW admissibility criterion applied to the asymptotic form of the bulk metric (BKSW). We first introduce the coordinate transformation \cite{Hawking:1998kw}:
\begin{align}
    \frac{\cos \vartheta}{z}=r \cos \theta ,
    \qquad
    \frac{1}{z^2}=\frac{r^2 \Delta_{\theta}+a^2 \sin^2\theta}{\Xi},
    \label{asymptotic-coordinates}
\end{align}
and then take the limit $r\to\infty$. The solution becomes
\begin{align}
    \diff s^2={}&\frac{\diff z^2}{z^2}+\frac{\beta^2\, \diff \tau^2+ \diff \vartheta^2 +\sin ^2\vartheta \, \left(\diff \phi-i \beta\,  \Omega \, \diff \tau \right)^2}{z^2}\,.
    \label{asymptotic-metric4d}
\end{align}
Since we have taken $\lim_{s \to \infty}\im \, r(s)= 0$, in this region the radial coordinate is real. The only non-trivial complex structure relevant for the KSW test is therefore contained in the $(\tau,\phi)$ block,
\begin{align}
   g_{(2)}= \left(
\begin{array}{cc}
 \frac{\beta ^2 \left(1-\Omega ^2 \sin ^2\vartheta \right)}{z^2} & -\frac{i \beta  \Omega  \sin ^2\vartheta}{z^2} \\
 -\frac{i \beta  \Omega  \sin ^2\vartheta}{z^2} & \frac{\sin ^2\vartheta}{z^2} \\
\end{array}
\right).
\label{two-dimensional-boundary-block}
\end{align}
Given
\begin{align}
    \det g_{(2)}=\frac{\beta ^2 \sin ^2\vartheta}{z^4},
\end{align}
the $p=0$ KSW condition gives precisely $\re \, \beta>0$, assuming $\sin\vartheta>0$.

Because the relevant block is two-dimensional, one also has to check the $p=1$ condition. This amounts to requiring the positive-definiteness of
\begin{align}
    A^{(2)}=\mathrm{Re}\left(\sqrt{\det\, g_{(2)}} g_{(2)}^{-1}\right)=
    \re \left(
\begin{array}{cc}
 \frac{\sin \vartheta}{\beta } & i \Omega  \sin \vartheta  \\
 i \Omega  \sin \vartheta &   \csc \vartheta\, \beta \left(1 - \Omega ^2 \sin^2 \vartheta \right) 
\end{array}
\right).
\label{Wmatrix}
\end{align}
We can use Sylvester's criterion to check for positive-definiteness of $A^{(2)}$, which amounts to checking that all leading principal minors ($A^{(2)}_{11}$ and $\det A^{(2)}$) of $A^{(2)}$ are positive. 
$A^{(2)}_{11} = \sin\vartheta\,\rm Re\,\beta^{-1}$ is automatically positive, as the condition $\rm Re\,\beta^{-1}>0$ is equivalent to $\rm Re\,\beta>0$. 

We then have to check that $\det A^{(2)}>0$. Writing this condition in terms of $\beta$ and $\omega$ we find
\begin{equation}
    \frac{\rm (\re \,\beta) ^2-\bigl(\rm \re\,\beta+\rm Re \,\omega\bigr)^2\sin^2\vartheta}{\rm (Re\,\beta)^2+(\rm Im\,\beta)^2}>0\,,
\end{equation}
which is strongest for $\sin \vartheta=1$, in which case it reduces to
\begin{equation}
    -\rm Re\,\omega\,\bigl(2\rm Re\,\beta+\re \,\omega\bigr) >0 \quad \longleftrightarrow \quad \frac{\bigl(\rho^2-(1-\sigma)^2\bigr)\bigl(1+\rho^2-\sigma^2\bigr)}{\rho^4+2\rho^2\bigl(\sigma^2-4\sigma+3\bigr)+\bigr(1-\sigma^2\bigl)^2}<0\,,
    \label{asymptKSWp1condition}
\end{equation} 
The condition~\eqref{asymptKSWp1condition}, together with $\rm Re\,\beta>0$, is more stringent than that imposed by the convergence of the index. It is the analogue of~\eqref{eq:nonBPSdescendantconvergence} and, as discussed in section~\ref{sec2}, it is required for the convergence of the partition function of the dual three-dimensional field theory which receives contributions also from non-BPS states. 
Within the horizon domain, however, this additional constraint is automatically satisfied. Consequently, the asymptotic KSW and the index-convergence conditions select the same allowed regions: 
\begin{align}
    \textbf{horizon domain}\,\cap\,\textbf{index convergence} \,=\, \textbf{horizon domain}\,\cap\,\textbf{BKSW}
\end{align}
Notice that, although the definition of the criterion formally requires considering all cases with $0\leq p\leq d$, the cases with $p>d/2$ do not provide independent constraints. As observed in \cite{Witten:2021nzp}, the condition associated with a $p$-form can be reduced to the corresponding condition for degree $d-p$. \footnote{For instance, in $d=2$, the criterion for the case $p=2$ reads $\re \left(\sqrt{\det g} \,F_{\mu \nu} F_{\rho \sigma}g^{\rho \mu}g^{ \sigma\nu}\right)=2\re \left(\sqrt{\det g} \,F_{12}^2 \left(g^{11}g^{22}-g^{12}g^{21}\right)\right)=2 \re\left(\frac{1}{\sqrt{\det g}} F_{12}^2\right)>0$. Therefore, the $p=2$ condition is equivalent to the $p=0$ condition, and it is sufficient to consider only $p=0$ and $p=1$ independently.}

\paragraph{Near-horizon KSW analysis.}
\label{subsubsec:outer-near-horizon}
In the near-horizon region, the local geometry takes the form~\eqref{eq:nh3} and depends on the tangent to the radial contour at the endpoint, $\kappa$. 
All metric components are now complex. Nevertheless, since the metric is diagonal in the coordinates $(R,\tau,\theta,\phi)$, the KSW condition can be tested using the form of the criterion given in~\eqref{ksw2}.

The dependence on $\kappa$ reflects the fact that the allowability of the near-horizon metric depends on the direction along which the horizon is approached in the complex plane. This direction can be chosen freely as it is associated with our freedom of choosing a specific contour. It follows that a given horizon is allowed if there exists at least one direction, i.e. one value of $\kappa$, for which the corresponding near-horizon metric is KSW-allowed. 

In order to check whether a given horizon is allowed by the near-horizon analysis one proceeds as follows. For each fixed value of $\kappa$ one first finds the region in the $\rho,\sigma$ parameter space for which the near-horizon geometry satisfies the KSW criterion for all values of $\theta$. One then repeats the analysis for different values of $\kappa$ selecting the union of all these allowed regions as $\kappa$ varies. This identifies the region in the $\rho,\sigma$ space containing horizons that can be reached along at least one direction. We refer to this procedure as the \emph{full} near-horizon KSW criterion. 

In practice, this procedure can be computationally heavy to implement for complicated metrics. A weaker version of the criterion can instead be used, allowing for an easier implementation and analytic control by still retaining a non-trivial necessary condition. 

To derive it, note that $\kappa$ multiplies only the $g_{RR}$ and $g_{\tau\tau}$ terms in the near-horizon metric~\eqref{eq:nh3}. The full near-horizon KSW criterion then reads
\begin{align}
    \text{Full near-horizon KSW}: \quad \,\exists\kappa\,,\,\, \text{s.t.}\,\,\,\, \forall \theta\,,\quad 2 \left|\mathrm{Arg} \,g_{RR}\right|&+\left|\mathrm{Arg} \,g_{\theta\theta}\right|+\left|\mathrm{Arg} \, g_{\phi\phi}\right| <\pi\,.
     \label{eq:FullNearHorizonCrit}
\end{align}
Now for each angle $\theta$ we choose $\kappa$ so as to cancel the phase in $g_{RR}$, and thus to drop out the $2 |\mathrm{Arg} \,g_{RR}|$ contribution in~\eqref{eq:FullNearHorizonCrit}. A similar procedure was used in \cite{Chen:2022hbi}. The constraint then reduces to
\begin{equation}
\text{Weak near-horizon KSW}: \qquad \forall\theta,\,\,\left|\mathrm{Arg} \,g_{\theta\theta}\right|+\left|\mathrm{Arg} \, g_{\phi\phi}\right| <\pi\,.
     \label{eq:WeakNearHorizonCrit}
\end{equation}
Importantly, the weaker constraint~\eqref{eq:WeakNearHorizonCrit} represents a necessary condition for the full constraint~\eqref{eq:FullNearHorizonCrit}. Indeed, if~\eqref{eq:WeakNearHorizonCrit} is not satisfied, then the contribution coming from $g_{\theta\theta}$ and $g_{\phi\phi}$ already violates the inequality for at least one angle. 
This result cannot be changed by any value of $\kappa$ as $g_{\theta\theta}$ and $g_{\phi\phi}$ are independent of it, and $|\mathrm{Arg} \,g_{RR}|$ can only further increase the lhs of~\eqref{eq:FullNearHorizonCrit}. 

The four-dimensional solution is simple enough that the full criterion~\eqref{eq:FullNearHorizonCrit} can be directly studied numerically. However, we find that the weak version is enough to provide new non-trivial results as compared to the boundary KSW criterion. In other cases, such as in the more involved bulk analysis or in the more complicated five-dimensional geometry, resorting to a weaker but necessary condition can be of great help.

\begin{figure}[htbp]
\centering
\includegraphics[width=0.6\textwidth]{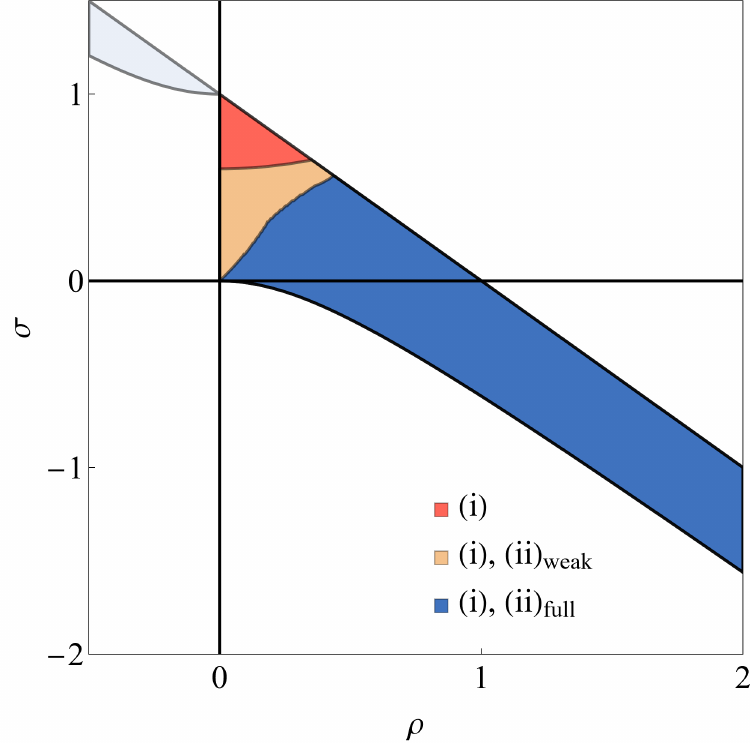}
\caption{Classification of the $(\rho,\sigma)$ parameter space belonging to the horizon domain. Each point is classified according to whether it (i) satisfies the convergence conditions of the superconformal index, (ii)$_{\rm{weak}}$ satisfies both the weak near-horizon condition~\eqref{sumargsreduced} and the asymptotic KSW criterion, (ii)$_{\rm{full}}$ satisfies both the full near-horizon condition and the asymptotic KSW criterion. The different regions correspond to the combinations of (i)-(ii) indicated in the legend.}
\label{fig:figure3}
\end{figure}

Let us then consider the weak version of the criterion~\eqref{eq:WeakNearHorizonCrit} for the outer horizon.  For $|a|<1$, it reduces to
\begin{equation}\label{sumargsreduced}
   \left|\mathrm{Arg} \left(a^2 \cos ^2\theta +r_+^2\right)\right|+\left|\mathrm{Arg} \left[ \frac{\left(a^2+r_+^2\right)^2 }{ \left(a^2 \cos ^2\theta +r_+^2\right)}\right]\right|<\pi\,.
\end{equation}
The left-hand side is non-decreasing for $\theta\in[0,\pi/2]$ and symmetric about $\theta=\pi/2$. The most restrictive condition is therefore attained at the equator and it is sufficient to evaluate~\eqref{sumargsreduced} at this angle. The region of the horizon domain in which the weak condition~\eqref{sumargsreduced} is satisfied is shown in figure~\ref{fig:figure3}.

Even resorting to the weaker version of the constraint, we obtain a non-trivial result. The near-horizon analysis excludes part of the horizon domain that satisfies the asymptotic KSW criterion (red region in figure~\ref{fig:figure3}). Note that the excluded region lies close to the imaginary axis, the $a=1$ line and the $\rho=0,\,\sigma=1$ singular point, thus removing (regular) solutions that lie too close to these singular black hole solutions.

Unlike the cases studied in \cite{BenettiGenolini:2025jwe,BenettiGenolini:2026raa,Krishna:2026rma}, KSW imposes restrictions beyond those required by index convergence and the KSW criterion in the boundary geometry.

We also tested the full version of the near-horizon KSW criterion~\eqref{eq:FullNearHorizonCrit} for various values of $\kappa$. In particular, the whole region with $\sigma <0$ is allowed by considering values of $\kappa$ with $\rm Arg\,\kappa <0$. Additional nontrivial constraints appear for $\sigma >0$. We tested the full near-horizon criterion for
$O(20)$ different values of the $\kappa$-parameter phase in the range $\rm{Arg}\,\kappa \in[0,0.5]$, with the resulting allowed region being the union of those found for each fixed value of $\kappa$. For $\rm{Arg}\,\kappa >0.5$  the near-horizon KSW criterion quickly becomes highly restrictive and no new allowed regions are included.
The largest region allowed by~\eqref{eq:FullNearHorizonCrit} is shown in figure~\ref{fig:figure3} (blue region). As compared to the weak version (blue plus yellow region in figure~\ref{fig:figure3}), the full criterion removes additional horizons that lie close to the singular boundaries of the outer horizon domain.

\paragraph{Bulk KSW analysis.}
We conclude our analysis by investigating the KSW criterion in the bulk geometry. Besides the conditions introduced in~\eqref{eq:ContourBoundaryConditions}, the contour in the radial complex plane must be such that the induced metric satisfies the KSW criterion at every point along the path and for every value of $\theta$.

A direct analysis of the full metric is technically involved. The bulk metric is generically non-diagonal and none of its components is guaranteed to be real. We therefore employ the third formulation of the KSW criterion, expressed directly in terms of the eigenvalues of the complex matrix $gA$. Moreover, the radial component depends explicitly on the tangent vector $r'(s)$. Consequently, the full KSW condition depends both on the position $r(s)$ in the complex plane, and on the direction along which the contour passes through that point. 

Similarly to what we just discussed in the near-horizon analysis, in order to make the problem more tractable we resort to a weaker version of the constraint.
For each fixed angle $\theta$ and at each bulk point, we fix $r'(s)$ so that $g_{ss}$ is real and positive. This allows us to discard the radial component $g_{ss}$ and study only the three-dimensional metric along the $(\tau,\theta,\phi)$ directions. By doing so we eliminate the dependence on $r'(s)$. The KSW criterion applied to the reduced metric components provides a necessary condition for the allowability of a given bulk point. The regions in the complex $r$-plane for which the above criterion is not satisfied are associated with points that
are necessarily incompatible with the full KSW criterion, for any possible choice of a contour, corresponding to different values of $r'(s)$.
This defines, in the complex $r$-plane, a set of ``forbidden'' bulk points.

We perform the analysis numerically on a uniform lattice in the complex $r$-plane. For each fixed horizon $r_+$, we consider a square region centered at $r_+$ and we check the weak criterion on each grid point.
The lattice points are  divided into an admissible set and a forbidden set. Since the metric also depends on $\theta$, we repeat the computation for the representative values of $\theta=\left(0,\frac{\pi}{4},\frac{\pi}{2}\right)$. A violation at any one of these values is sufficient to exclude the corresponding point. 

\begin{figure}[htbp]
\centering
\begin{subfigure}[t]{0.49\textwidth}
    \centering
    \includegraphics[width=\textwidth]{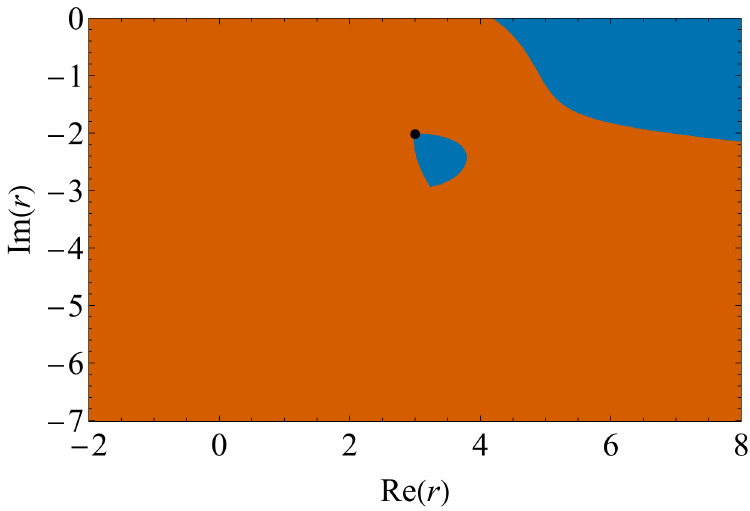}
    \caption{Bulk analysis for $(\rho,\sigma)=(3,-2.02)$.}
    \label{fig:bulk-first}
\end{subfigure}
\hfill
\begin{subfigure}[t]{0.49\textwidth}
    \centering
    \includegraphics[width=\textwidth]{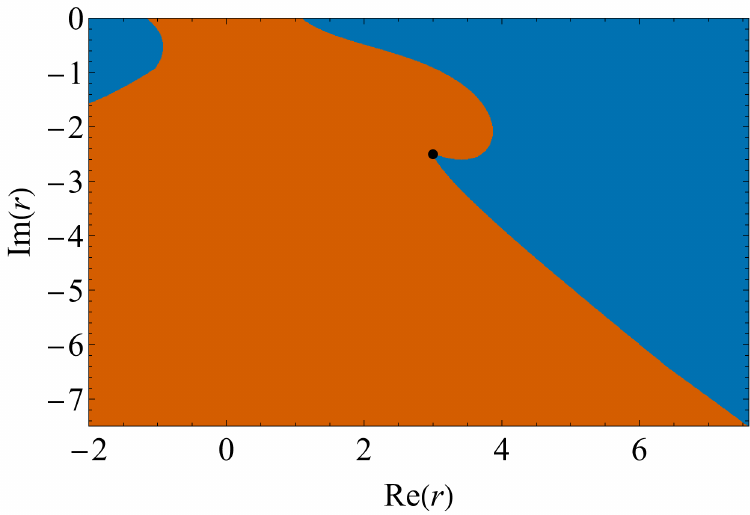}
    \caption{Bulk analysis for $(\rho,\sigma)=(3,-2.5)$}
    \label{fig:bulk-second}
\end{subfigure}

\caption{Numerical bulk analysis of the reduced metric, for fixed \(r_+=\rho+i\sigma\) and \(\theta=\pi/2\). Blue points satisfy the reduced KSW conditions, whereas orange points violate them. In the left panel there is no KSW-allowed complex contour for the $r$-coordinate.}
\label{fig:figure4}
\end{figure}

Again, even resorting to the weaker version of the criterion allows us to obtain new non-trivial results:
even when the KSW criterion is satisfied in a neighborhood of the horizon and in the asymptotic region, the corresponding admissible domains may be separated by an intermediate forbidden region. In such a case, no
continuous contour $r(s)$ is allowed to connect the horizon to the asymptotic boundary without crossing a region in which the reduced metric necessarily violates the criterion. An example of this obstruction is shown in figure~\ref{fig:figure4}.

\begin{figure}[htbp]
\centering
\includegraphics[width=0.57\textwidth]{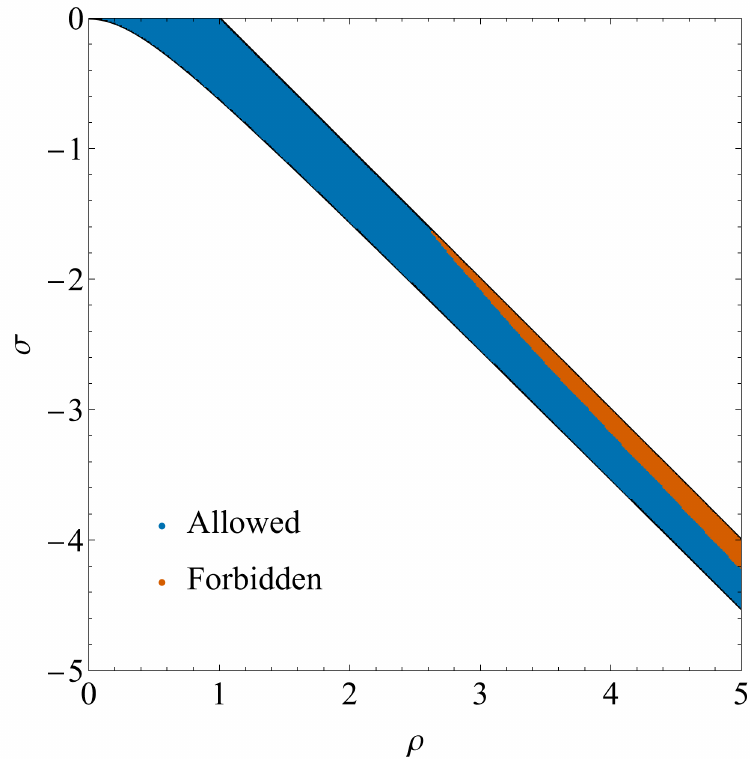}
\caption{Numerical classification of complex horizons in the $(\rho,\sigma)$-plane obtained from the reduced bulk analysis. Allowed points may still be removed by a full bulk analysis.}
\label{fig:figure5}
\end{figure}

For each horizon, we construct the corresponding grid in the complex $r$-plane, determine the points that satisfy the reduced KSW criterion, and test whether the near-horizon admissible region is separated from the asymptotic region. The classification for $\sigma<0$ is shown in figure~\ref{fig:figure5}; for $\sigma>0$, the bulk analysis imposes no additional exclusions beyond the near-horizon analysis.
We check this for $\mathcal{O}(10^4)$ points in the $(\rho,\sigma)$-plane, corresponding to different values of the complex horizon parameter $r_+$.

The KSW criterion in the bulk imposes additional constraints on the parameter space. 
Interestingly, it removes a previously allowed region that lies close to the upper boundary of the horizon domain. This boundary is related to singular solutions with $a=1$ and $\re\,\omega =0$. Therefore, as in the near-horizon analysis, the bulk KSW criterion removes (regular) solutions that lie too close to the edges of the horizons domain. As $\rho\to+\infty$, the forbidden region progressively widens, approaching the lower boundary of the horizon domain, corresponding to solutions with $a=0$.

\begin{figure}[htbp]
\centering
\begin{subfigure}[t]{0.48\textwidth}
    \centering
    \includegraphics[width=\textwidth]{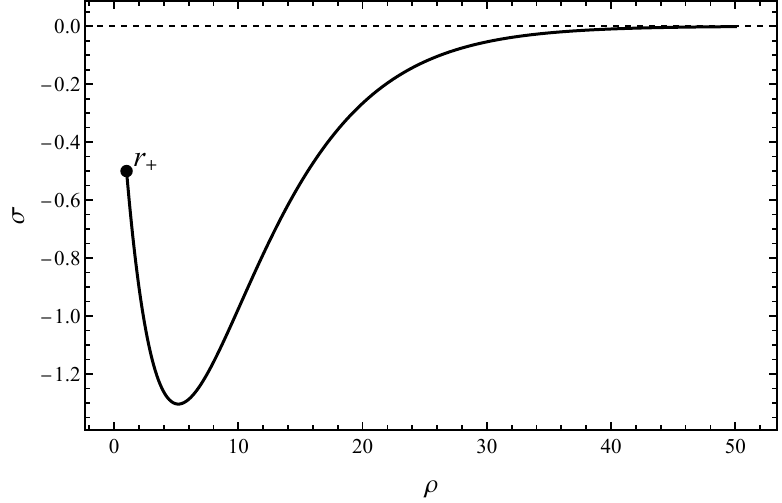}
    \caption{Profile of the complex radial contour in the \(r\)-plane.}
    \label{fig:first}
\end{subfigure}
\hfill
\begin{subfigure}[t]{0.48\textwidth}
    \centering
    \includegraphics[width=\textwidth]{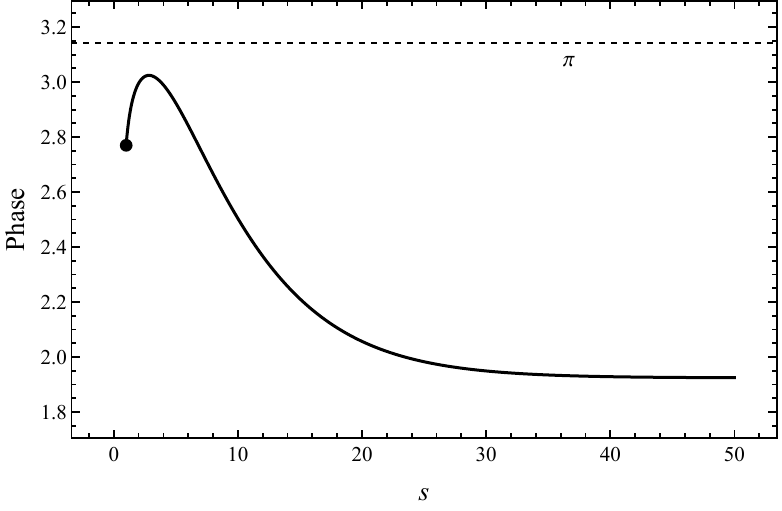}
    \caption{\(\sum_i \left|\arg \mu_i(gA)\right|\), evaluated along the contour at \(\theta=\pi/2\).}
    \label{fig:second}
\end{subfigure}

\caption{Explicit example of a complex radial contour for which the metric satisfies the KSW criterion throughout the bulk. The contour is parametrized as
$
r(s)=s+i\left[
(s-\rho)\left(\frac{\sigma}{\lambda}-\gamma\right)+\sigma
\right] \rme^{-\frac{s-\rho}{\lambda}},
$
with $(\rho,\sigma,\lambda,\gamma)=(1,-0.5,5,0.5)$.} 
\label{fig:combined}
\end{figure}

Our analysis does not yet establish the existence of an admissible contour for every point that survives the reduced test. A definitive verification of admissibility requires constructing an explicit contour $r(s)$ in the complex plane and considering the full metric at every point $(s,\theta)$, including the contribution proportional to $r'(s)^2$. In figure~\ref{fig:combined}, we provide an explicit example of such a contour for a specific value of $r_+$. We found this contour by trial and error and checked its existence for $\mathcal{O}(100)$ points.

\section{Charged, rotating AdS\texorpdfstring{$_5$}{} black hole}\label{sec:5d}
We now turn to the charged, rotating AdS$_5$ black hole solution and to its supersymmetric extension obtained by complexifying the $r_+$ parameter. We first describe the structure of the resulting complex horizons and their thermodynamic properties. We then analyze their KSW allowability and compare the resulting constraints with those required for convergence of the $\mathcal{N}=1$ SCI of the dual field theory.

\subsection{General charged, rotating solution}
The bosonic sector of $D=5$ pure gauged supergravity is given by\footnote{We use the normalization of \cite{Cabo-Bizet:2018ehj} for the gauge field and the corresponding electric charge. These are related to the ones in~\cite{Chong:2005hr} as $A^{\rm CCLP} = \frac{2\ell}{\sqrt{3}} A$, $\,Q^{\rm CCLP} = \frac{\sqrt{3}}{2\ell}Q\,$.} 
\begin{equation}
    S \,=\, \frac{1}{16\pi}\int\,\diff^5x\,\sqrt{-g}\left(R+\frac{12}{\ell^2}-\frac{\ell^2}{3}F^2-\frac{2\ell^3}{27}\epsilon^{\mu\nu\rho\sigma\lambda}F_{\mu\nu}F_{\rho\sigma}A_\lambda \right)\,,
\end{equation}\label{eq:metric}
where $\ell$ is the radius of the AdS$_5$ solution and we set $G=1$. This theory admits a charged, doubly-spinning, asymptotically AdS$_5$ black hole solution found in \cite{Chong:2005hr}. In a frame that is co-rotating with the horizon, the Euclidean metric reads (setting $\ell=1$)
\begin{align}\label{CCLPmetric}
\nonumber \diff s^2\,&=\,\frac{\beta  \Delta_\theta   \left[\beta  \Sigma ^2 \left(1+r^2\right) \,\diff \tau+2 i \nu  q\right]\diff \tau}{\Xi_a \Xi_b\, \Sigma ^2}+\frac{2 \nu  q \omega }{\Sigma ^2}+\frac{\left(r^2+a^2\right)}{\Xi_a}\, \sin^2\theta\,(\diff \phi-i \beta  \Omega_1 \,\diff \tau)^2\\
\nonumber &\quad+\frac{\left(r^2+b^2\right)}{\Xi_b}\, \cos ^2\theta  \,(\diff \psi-i \beta  \Omega_2 \,\diff\tau )^2+\Sigma ^2 \left(\frac{\diff r^2}{\Delta_r}+\frac{\diff\theta^2}{\Delta_\theta }\right)+\frac{f}{\Sigma ^4} \left(\omega +\frac{i \beta  \Delta_\theta  \,\diff\tau }{\Xi_a \Xi_b}\right)^2\,,\\
A\,&=\,-\frac{3q}{2\Sigma^2}\left(\frac{i \beta\Delta_\theta\, \diff \tau}{\Xi_a \Xi_b}+\omega\right)+i \beta \Phi \,\diff \tau\,,
\end{align}
and
\begin{align}\label{eq:defs}
\nonumber
\nu &= a \cos^2\theta \left(\diff \psi-i \beta \Omega_2 \,\diff \tau \right)+b \sin^2\theta \left(\diff \phi-i \beta \Omega_1 \,\diff \tau \right)\,,
\\
\nonumber
\omega &=\frac{a \sin^2\theta}{\Xi_a}\left(\diff \phi-i \beta \Omega_1 \,\diff \tau \right)+\frac{b \cos^2\theta}{\Xi_b}\left(\diff \psi-i \beta \Omega_2 \,\diff \tau \right)\,,
\\
\nonumber
\Delta_\theta &=1-a^2\cos^2\theta-b^2\sin^2\theta, \qquad \qquad
\Xi_a =1-a^2\,,
\qquad \qquad
\Xi_b=1-b^2\,,
\\
\Delta_r &=\frac{\left(r^2+1\right)\left(r^2+a^2\right)\left(r^2+b^2\right)+2abq+q^2}{r^2}-2m\,,
\\
\nonumber
\Sigma^2 &=r^2+a^2\cos^2\theta+b^2\sin^2\theta\,,\qquad\qquad f=2m\Sigma^2+2abq\Sigma^2-q^2\,, 
\end{align}
while $\beta,\Omega_1,\Omega_2,\Phi$ are the chemical potentials, to be defined below.

The solution is parametrized by the coordinates $(\tau,r,\theta,\phi,\psi)$ with $\tau \sim \tau+1$ and angular coordinates $\phi,\psi$ taken $2\pi$ periodic, i.e. they satisfy untwisted periodicity conditions. The angular coordinate $\theta$ takes values in $[0,\pi/2]$. The solution depends on four parameters $m,q,a,b$, with $a^2<1$, $b^2<1$ in order for $\Xi_a,\Xi_b$ and $\Delta\theta$ to be non-vanishing for all values of $\theta$.
Correspondingly, it carries four conserved quantities: mass $M$, electric charge $Q$, and angular momenta $J_1,J_2$:
\begin{align}\label{eq:5D2J1QCharges}
    M &= \frac{\pi m(2\Xi_a+2\Xi_b-\Xi_a\Xi_b)+2\pi abq(\Xi_a+\Xi_b)}{4\Xi_a^2\,\Xi_b^2}\,, \qquad Q=\frac{\pi\,q}{2\Xi_a\,\Xi_b}\,,\notag \\[2mm]
    J_1 &= \frac{\pi \bigl[2am+qb(1+a^2)\bigr]}{4\Xi_a^2\,\Xi_b}\,, \qquad
    J_2 = \frac{\pi \bigl[2bm+qa(1+b^2)\bigr]}{4\Xi_a\,\Xi_b^2}\,.
\end{align}
As before, the metric admits multiple complex horizons, located at the roots $r_i,\,i=1,\cdots,6$ of the sextic polynomial $r^2\Delta_r(r)$. As this only depends on $r^2$, its roots are organized in pairs $(r_i,-r_i)$. Equivalently we can regard it as a cubic polynomial in $r^2$ and work with the three roots $r_i^2$,
\begin{equation}
    \Delta_r(r) \,=\, \frac{1}{r^2}\prod_{i=1}^{3}(r^2-r_i^2)\,.
\end{equation}
Comparing with the expression for $\Delta_r$ in~\eqref{eq:defs}, we get the relations:
\begin{equation}
    \begin{cases}
        \sum_{i=1}^{3}r_i^2 = -(a^2+b^2+1)\,,\\[1mm]
        \sum_{i<j} r_i^2r_j^2 = (a^2+b^2-2m)+a^2b^2\,,\\[1mm]
        r_1^2r_2^2r_3^2 = -(ab+q)^2\,.
    \end{cases}
    \label{eq:5D2J1QRadiiSymmRelations}
    \qquad 
\end{equation}
 From either the first or the last equation in~\eqref{eq:5D2J1QRadiiSymmRelations}, it follows at least one of the three roots $r_i^2$ must be negative, and is therefore associated with a virtual horizon, which in this case is purely imaginary.

The remaining two roots $r_i^2$ are either both real and positive, corresponding to the outer event horizon $r_+$ and the inner Cauchy horizon $r_0$, both real and negative, or a complex conjugate pair. As before, the last case will be relevant for the supersymmetric solutions we consider.

Associated with the real event horizon is the Bekenstein-Hawking entropy:
\begin{equation}
    S = \frac{\pi^2 \left[(r_+^2+a^2)(r_+^2+b^2)+abq\right]}{2\Xi_a\,\Xi_b\,r_+}\,,
    \label{eq:5D2J1QEntropy}
\end{equation}
and the chemical potentials:
\begin{align}
    T&= \frac{r_+^4\left(1+2r_+^2+a^2+b^2\right)-(ab+q)^2}{2\pi\,r_+ \left[(r_+^2+a^2)(r_+^2+b^2)+abq\right]}\,, \quad\ \Phi = \frac{3\,q\,r_+^2}{2\left[(r_+^2+a^2)(r_+^2+b^2)+abq\right]}\,,\notag\\[3mm] 
    \qquad\Omega_1 &= \frac{a(r_+^2+b^2)(1+r_+^2)+bq}{(r_+^2+a^2)(r_+^2+b^2)+abq}\,,\qquad \Omega_2 = \frac{b(r_+^2+a^2)(1+r_+^2)+aq}{(r_+^2+a^2)(r_+^2+b^2)+abq}\,.
    \label{eq:5D2J1QChemPot}
\end{align}
that satisfy the first law of thermodynamics and the quantum statistical relation 
\begin{align}
    \diff M &=T\,\diff S+\Omega_1\,\diff J_1+\Omega_2\, \diff J_2+\Phi\, \diff Q\,, \notag \\
    I &= \beta\,M-S-\beta\,\Omega_1\,J_1-\beta\,\Omega_2\,J_2-\beta\,\Phi\,Q\,.
\end{align}
Similar thermodynamic quantities can be associated with the other horizons by replacing the variable $r_+$ appearing  in eqs.~\eqref{eq:5D2J1QEntropy} and~\eqref{eq:5D2J1QChemPot} with the corresponding horizon radius $r_i$.

\subsection{Supersymmetry and complexification}
The condition for the above solution to preserve supersymmetry reads \cite{Chong:2005hr}
\begin{equation}
q = \frac{m}{1+a+b}\,.
\label{eq:5dSusyParameters}
\end{equation}
Imposing $\Delta_r(r_i)=0$ allows us to trade the parameter $m$ for a given horizon radius $r_i$
\begin{equation}
    m = -\bigl(1+a+b\bigr)\bigl(1\pm ir_i)\bigl(a\pm ir_i\bigr)\bigl(b\pm  ir_i\bigr)\,.
    \label{eq:5dmass-root-relation}
\end{equation}
As before, the BPS solution is obtained by requiring that $a,b,m \in \mathbb R$ and that at least one real positive root $r_+>0$ of $\Delta_r$ exists. This requirement is satisfied if 
\begin{equation}
\begin{cases}
    m =(a+b)(1+a)(1+b)(1+a+b)\,, \\[2mm]
    a+b+ab>0\,,
    \end{cases}
\end{equation}
in which case the roots of $\Delta_r$ are given by
\begin{equation}
   r_+^2 = r^{\star\,2} = a+b+ab\,, \qquad r_-^{\star2} = -(1+a+b)^2\,,
\end{equation}
with $r^{\star 2}$ being a double root and $r^\star_-$ being purely imaginary as expected.

Outside of the BPS locus, i.e. imposing just~\eqref{eq:5dSusyParameters}, the chemical potentials read
\begin{align}
    \beta &= \mp2\pi i\,\frac{ r_i^2\,(a\pm i r_i)(b\pm ir_i)(r^{\star 2}\mp i r_i)}{r_i^4\bigl(2r_i^2+1+a^2+b^2\bigr)-\bigl(a b-(a\pm ir_i)(b\pm ir_i)(1\pm ir_i)\bigr)^2}\,, \notag \\[2mm]
    \Omega_1 &= \frac{(1\pm ir_i)(r^{\star\,2}\mp ia\, r_i)}{(a\pm i r_i)(r^{\star 2}\mp i r_i)}\,, \qquad  \Omega_2 = \frac{(1\pm ir_i)(r^{\star\,2}\mp ib\, r_i)}{(b\pm i r_i)(r^{\star 2}\mp i r_i)}\,, \qquad \Phi= \mp\frac{3i r_i(1\pm i r_i)}{2(r^{\star\,2}\mp i r_i)}\,,
    \label{eq:5dsusychempot}
\end{align}
while the associated field theory variables are given by
\begin{align}\label{chempot}
\nonumber \omega_1 \,&=\, \beta \left(\Omega_1-1\right) \,=\,
\pm\,\frac{2\pi i(a - 1)(b \pm i r_i)}
{({r^*}^{2} - 3 r_i^{2})\pm2ir_i(1 + a + b)} \,, \\[6pt]
\omega_2 \,&=\,\beta \left(\Omega_2-1\right) \,=\,
\pm\frac{2\pi i(b - 1)(a \pm i r_i)}
{({r^*}^{2} - 3 r_i^{2})\pm2ir_i(1 + a + b)} \,, \\[6pt]
\nonumber \varphi \,&=\, \beta \left(\Phi-\frac{3}{2}   \right) \,=\,
\pm\frac{3\pi i\,(a \pm i r_i)(b \pm i r_i)}
{({r^*}^{2} - 3 r_i^{2})\pm 2ir_i(1 + a + b)} \,,
\end{align}
and satisfy the supersymmetry constraint
\begin{align}\label{susyconstraint}
    \beta\left(1+\Omega_1+\Omega_2-2\Phi\right) = \omega_1+\omega_2-2\varphi=\mp2\pi i\,.
\end{align}

\paragraph{Supersymmetric solutions with real charges.}
As in the four-dimensional case, we consider supersymmetric solutions with real conserved charges. For this choice, at least one of the three roots $r_i^2$ of $\Delta_r$ is negative, corresponding to a purely imaginary virtual horizon $r_-$. The remaining two roots are either both negative or form a complex conjugate pair, $(r_+^2,\bar r_+^2)$.

Writing $r_\pm = \rho_\pm+i\sigma_\pm$ in the expression for $m=m(r_\pm)$ in~\eqref{eq:5dmass-root-relation} one finds:
\begin{equation}
    \text{Im} \,m(r_\pm) =-(1+a+b)\,\rho_\pm\,\Bigl(r^{\star 2}-\rho_\pm^2+\sigma_\pm\bigl(3\sigma_\pm-2(1+a+b)\bigr)\Bigl)\,,
    \label{eq:5dImmparameter}
\end{equation}
where we considered the upper sign choice in~\eqref{eq:5dmass-root-relation} both for $r_+$ and $r_-$. For the related roots $-r_i$ and $\pm\bar r_i$, we must consider a sign choice consistent with the one made for $r_i$. In particular, from the expression of $\Delta(r)$ in~\eqref{eq:defs} it follows that $m(r_i,a,b)=m(-r_i,a,b)$ and $m(r_i,a,b) = \bar m(r_i,a,b)$. Therefore, once a sign choice is made for $r_i$, consistency requires us to consider the opposite one for $-r_i$ and $\bar r_i$, and the same for $-\bar r_i$. Then,
up to an overall $-1$ factor,~\eqref{eq:5dImmparameter} holds for the four roots $\pm r_i$ and $\pm \bar r_i$ written in terms of $\rho$ and $\sigma$. 

Imposing the reality condition $\im\,m=0$ yields two possible solutions. The branch associated with the complexified outer horizon is
\begin{equation}
    r_+:\, \begin{cases}
        \rho_+ = \pm\sqrt{r^{\star\,2} +\sigma_+\bigl(3\sigma_+-2(1+a+b)\bigr)} \,,\\[1mm]
        m=(1+a+b)(1+a-2\sigma_+)(1+b-2\sigma_+)(a+b-2\sigma_+)\,,
    \end{cases}
    \label{eq:5dComplexRplus}
\end{equation}
where we define $r_+$ as the root with Re $r_+>0$ so that, in the BPS limit $\sigma_+ \to0$, one recovers a real event horizon. The other branch is instead related to $-r_+$.

The solution associated with the virtual horizon reads
\begin{equation}
    r_-:\, \begin{cases}
        \rho_- = 0\,, \\[1mm]
        m=(1+a+b)(\sigma_--a)(\sigma_--b)(\sigma_--1)\,.
    \end{cases}
\end{equation}
Furthermore, requiring that $\rho_+ \in \mathbb R$ together with the constraints $a^2<1$ and $b^2<1$ fixes the range of $\sigma_+$ to lie in one of two branches defined by
\begin{align}
   \text{Branch 1:}\qquad \sigma_+ &\le\frac{1}{3}\left(1+a+b-\sqrt{1+a^2+b^2-r^{\star\,2}}\right)\,, \notag \\[2mm]
   \text{Branch 2:}\qquad \sigma_+&\ge\frac{1}{3}\left(1+a+b+\sqrt{1+a^2+b^2-r^{\star\,2}}\right)\,.
\end{align}
The argument of the square root is always positive for $a^2<1$ and $b^2<1$. Notice that only branch 1 is continuously connected with the BPS solutions, which are recovered at $\sigma_+=0$. By contrast, branch 2 always satisfies $\sigma_+>0$ and therefore cannot be connected to a BPS configuration. When $\sigma_+$ saturates the bounds above, one has $\rho_+=0$. 
We have not imposed the additional condition $a+b+ab>0$, which is required for the regularity of the BPS solutions but is not needed for the supersymmetric non-extremal configurations considered here. This is to be contrasted with the conditions $a^2<1$ and $b^2<1$, which remain necessary to ensure regularity of the metric.

\begin{figure}[htbp]
\centering
\includegraphics[width=1\textwidth]{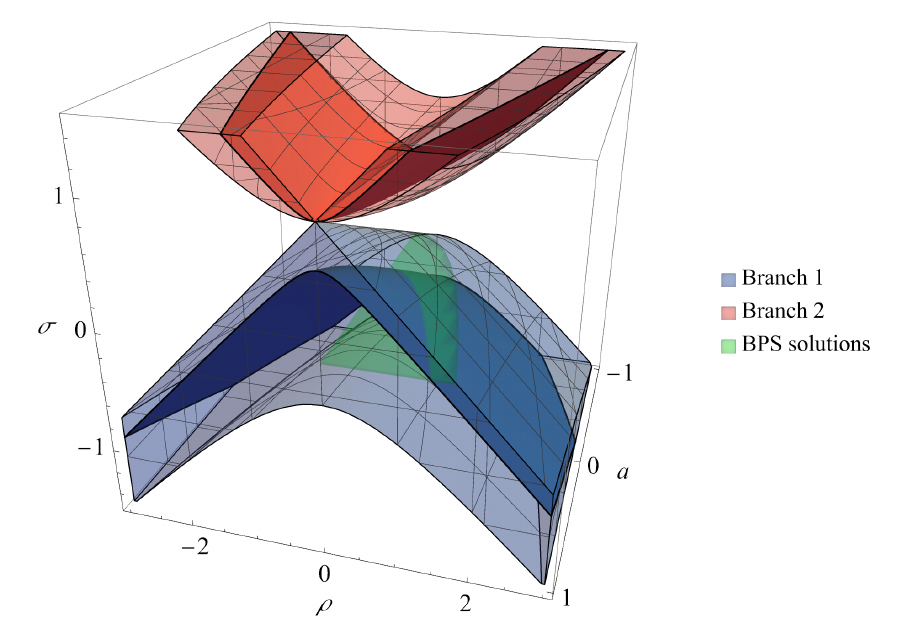}
\caption{Regions obeying the regularity constraints $a^2<1$ and $b^2<1$ in the $\rho,\sigma,a$ space. As commented in the text, only the first branch is connected with the BPS solutions, found at $\sigma=0$. The solid surfaces contained in the allowed regions provide a representative example of the $b\propto a$ sections. Here, the case $b=a/2$ is shown.}
\label{fig:5dAllowedRegion_arhosigma}
\end{figure}

Alternatively, for the outer horizon $r_+$ we may use $\rho$, $\sigma$, and $a$ as independent parameters, thereby adopting a parametrization  analogous to that used in the four-dimensional case. In terms of these variables, $b$ and $m$ are given by
\begin{align}
    b &= \frac{\rho_+^2-a+\sigma_+\bigl(2(a+1)-3\sigma_+\bigl)
    }{1+a-2\sigma_+}\,,\notag \\[1mm] 
    m &=\frac{\bigl(\rho_+^2+(a-\sigma_+)^2\bigr)\bigr(\rho_+^2+(1-\sigma_+)^2\bigl)\bigl(1+a+a^2+\rho_+^2-3\sigma_+^2\bigr)}{(1+a-2\sigma_+)^2}\,.
    \label{eq:5dbmtoARhoSigma}
\end{align}
In this parametrization, the constraints $a^2<1$ and $b^2<1$ defining the two branches become
\begin{align}
    \text{Branch 1}:&\qquad a-\sqrt{3+a^2+3\rho_+^2}\,<\,3\sigma_+\,<\,2+a-\sqrt{3\rho_+^2+(1-a)^2}\,, \notag \\[2mm]
    \text{Branch 2}:&\qquad a+\sqrt{3+a^2+3\rho_+^2}\,<\,3\sigma_+\,<\,2+a+\sqrt{3\rho_+^2+(1-a)^2}\,.
    \label{eq:5dBranches_arhosigma}
\end{align}
As before, the same parametrization extends to the three roots related to $r_+$ by reflection and complex conjugation. The regions of the $(\rho,\sigma,a)$ parameter space associated with the two branches in~\eqref{eq:5dBranches_arhosigma} are shown in figure~\ref{fig:5dAllowedRegion_arhosigma}.

The chemical potentials $\omega_{1,2}$ associated with the outer horizon are given by
\begin{align}
\omega_{1+} &= \frac{\pi(1-a)\bigl(i\rho_+-(a-\sigma_+)\bigr)\bigl(i\rho_++\sigma_+-1\bigr)}{\rho_+\Bigl(\bigl(1+a-2\sigma_+\big)^2-\bigl(i\rho_+-(a-\sigma_+)\bigr)\bigl(i\rho_++\sigma_+-1\bigr)\Bigr)} \,, \notag \\[2mm]
\omega_{2+} &= \frac{\pi(i\rho_+-\sigma_++a)\bigl(\rho_+^2+(\sigma_+-a)^2-(1+a-2\sigma_+)^2\bigr)}{\rho_+\Bigl(\bigl(1+a-2\sigma_+\big)^2-\bigl(i\rho_+-(a-\sigma_+)\bigr)\bigl(i\rho_++\sigma_+-1\bigr)\Bigr)} \,,
\label{eq:5dSusyChemPotrhosigma}
\end{align}
the expression for $\beta_+$ is more complicated and will not be shown here. $\varphi_+$ can be obtained by using~\eqref{susyconstraint}.

Defining $\omega_{1+}^\star(\rho_+,a)\equiv \omega_{1+}(\rho_+,a,\sigma_+=0)$, and similarly for $\omega_{2+}^\star$ and $\varphi_+^\star$, as the corresponding BPS values of the chemical potentials, one finds:
\begin{align}
\nonumber    \omega_{1+}(\rho_+,a,\sigma_+) &= \omega_{1+}^\star \left(\frac{\rho_+}{1-\sigma_+},1-\frac{1-a}{1-\sigma_+}\right)\,, \\[2mm]
     \omega_{2+}(\rho_+,a,\sigma_+) &= \omega_{2+}^\star \left(\frac{\rho_+}{1-\sigma_+},1-\frac{1-a}{1-\sigma_+}\right)\,.  
\end{align}
Also in this case, the supersymmetric chemical potentials only take values on the BPS locus, which is now parametrized by $\rho_+$ and $a$. We provide a derivation of this result from the extremization principle and assuming real charges, in the $J_1=J_2$ case, in appendix~\ref{app:RealChargExtrPrinc}.

For virtual horizons $r_-$, the inverse temperature and the chemical potentials are given in terms of $a$, $b$, and $\sigma_-$ by
\begin{align}
   \notag \beta_- &= -\frac{2\pi i\,(a-\sigma_-)(b-\sigma_-)(r^{\star\, 2}+\sigma_-)}{(r^{\star \,2}+\sigma_-^2)\bigl[r^{\star \,2}+\sigma_-\bigl(3\sigma_--2(1+a+b)\bigr)\bigr]}\,, \\[1mm]
    \omega_{1-} &=-\frac{2\pi i(1-a)(b-\sigma_-)}{r^{\star \,2}+\sigma_-\bigl(3\sigma_--2(1+a+b)\bigr)}\,, \\[1mm]\notag
    \omega_{2-}&= -\frac{2\pi i(1-b)(a-\sigma_-)}{r^{\star \,2}+\sigma_-\bigl(3\sigma_--2(1+a+b)\bigr)}\,.
\end{align}
These quantities are purely imaginary throughout the allowed parameter space. More generally, from the expressions of the supersymmetric chemical potentials in~\eqref{eq:5dsusychempot} and~\eqref{chempot}, a horizon associated with a purely imaginary root $r_i$ has $\beta\in i\mathbb{R}$, while $\Omega_{1,2}$ and $\Phi$ are real. Consequently, the corresponding field-theory chemical potentials $\omega_{1,2}$ and $\varphi$ are purely imaginary. Therefore, the index convergence conditions are not satisfied.

Since $\beta$ is purely imaginary, the asymptotic metric (given in eq.~\eqref{asymptotic-metric}) has Lorentzian rather than Euclidean signature. Indeed, writing $\beta=i|\beta|$, the ``Wick rotation''  $t=-i\beta \tau$ becomes a rescaling of the Lorentzian time coordinate $t =-i\beta \tau = |\beta|\tau$. However, in the near-horizon region, the coordinate that plays the role of an angular coordinate in parametrizing the cigar geometry is still given by $\tau$. It follows that $\tau$ should be periodically identified as usual. Since $\tau$ is a timelike rather than a Euclidean coordinate at the boundary, this identification necessarily introduces closed timelike curves in the boundary region and in particular in the background where the dual CFT lives. We take this result as a further indication that the saddles associated with the virtual horizons should not be included.

\subsection{Allowability of the complex solution}
We now study the KSW allowability of the complex solution, focusing on radial contours terminating at an outer horizon. In order to reduce the number of independent parameters, we restrict our analysis to the four representative cases

\begin{align}\label{cases}
b=a,\qquad b=\frac{a}{2},\qquad b=0,\qquad b=-a,
\end{align}
thereby eliminating one real parameter. For each of these cases, we use~\eqref{eq:5dbmtoARhoSigma} to express $a$ and $m$ in terms of the horizon variables $\rho_+$ and $\sigma_+$, and henceforth omit the $+$ subscripts. Note that the $b=-a$ sections of the full parameter space shown in figure~\ref{fig:5dAllowedRegion_arhosigma} do not intersect the BPS solutions, as they do not satisfy the bound $a+b+ab>0$. This is not an issue, as a solution in the first branch with $a=-b$ can still be connected to any given BPS solution via appropriate paths in the horizon domain shown in figure~\ref{fig:5dAllowedRegion_arhosigma}. In any case, independently of whether a supersymmetric solution can be connected with a given BPS solution, it still provides a genuine saddle of the GPI. For instance, this is the case for any virtual horizon, which can never be associated with a regular BPS saddle.
 
\begin{figure}[ht!]
\begin{subfigure}{0.48\textwidth}
    \centering
    \includegraphics[width=\textwidth]{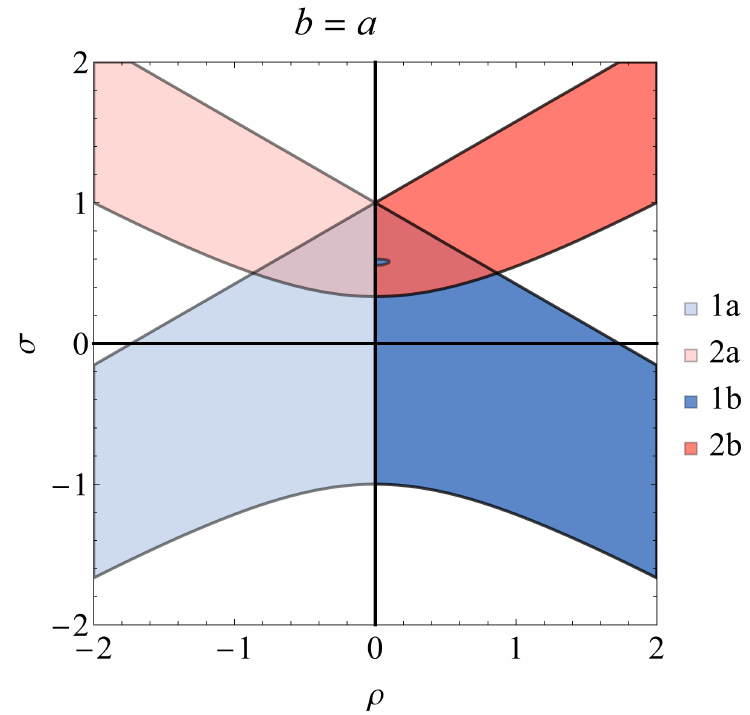}
\end{subfigure}
\hfill
\begin{subfigure}{0.48\textwidth}
    \centering
    \includegraphics[width=\textwidth]{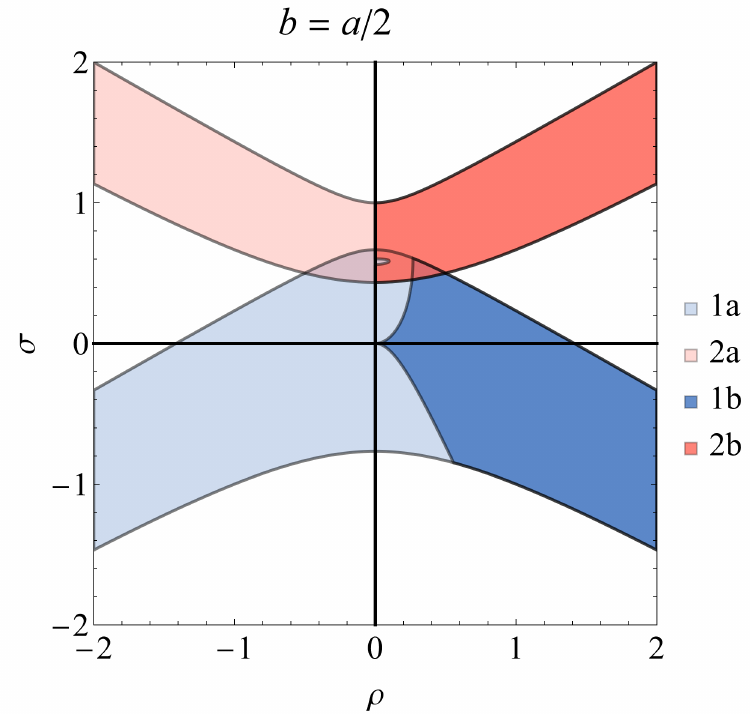}
\end{subfigure}

\begin{subfigure}{0.48\textwidth}
    \centering
    \includegraphics[width=\textwidth]{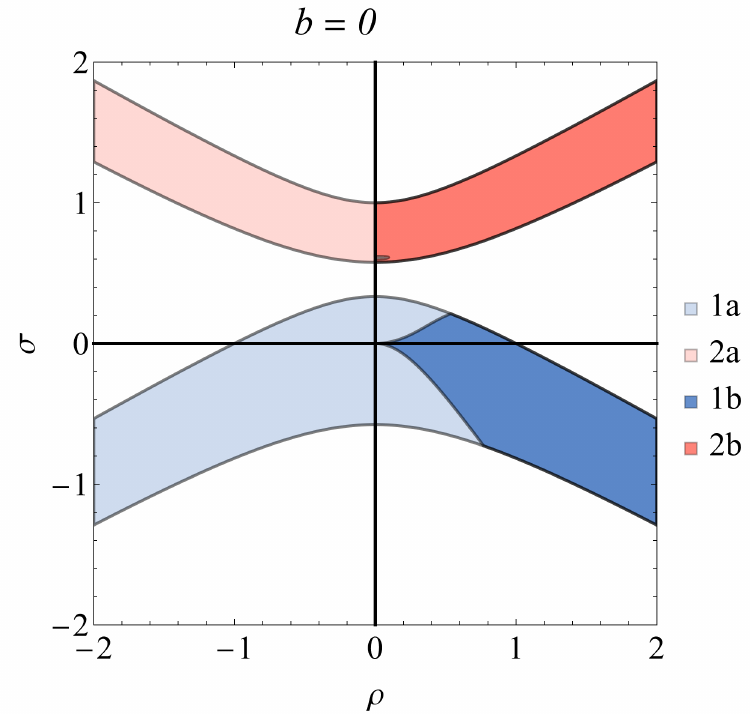}
\end{subfigure}
\hfill
\begin{subfigure}{0.48\textwidth}
    \centering
    \includegraphics[width=\textwidth]{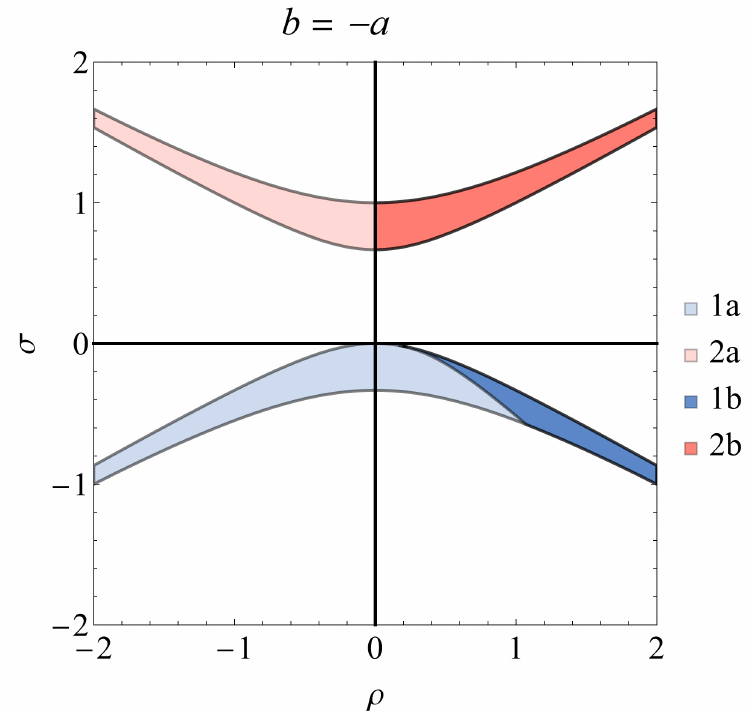}
\end{subfigure}

\caption{Classification of the $(\rho,\sigma)$ parameter space for $b\in\{a,\,a/2,\,0,\,-a\}$, including both outer horizon branches. In the legend, (1a) and (2a) denote the first- and second-branch outer horizon domains, respectively, while (1b) and (2b) denote the corresponding subdomains satisfying the convergence conditions of the SCI.}
\label{fig:figure7}
\end{figure}

\paragraph{Boundary convergence.} The convergence of the field theory index requires:
\begin{align}\label{indexconvergence5d}
    \re \,\beta>0\,, \qquad \qquad \re\, \omega_1<0\,, \qquad \qquad \re\, \omega_2<0\,.
\end{align}
For each of the choices in~\eqref{cases}, we display in figure~\ref{fig:figure7} the domains corresponding to the two branches in the complex horizon plane, together with the subsets that also satisfy the conditions~\eqref{indexconvergence5d}. The condition $\re\,\beta>0$ selects the half-plane $\rho>0$ (at least for the first branch). This is similar to what happens in four dimensions where every allowed horizon has $\rho>0$. For $a \neq b$ the last two conditions in~\eqref{indexconvergence5d} further remove, in the first branch, a region close to the $\rho =0$ axis. 

\paragraph{Asymptotic KSW analysis.}
We now consider the KSW criterion in the asymptotic bulk region. As previously discussed, we are interested in comparing the resulting constraints with the index-convergence conditions.
We concentrate on the first branch of solutions for more clarity in the visualization of our results. The asymptotic analysis does not exclude the second branch, which however is fully removed once the KSW criterion in the near-horizon region is considered.

\begin{figure}[htbp]
\begin{subfigure}{0.48\textwidth}
    \centering
    \includegraphics[width=\textwidth]{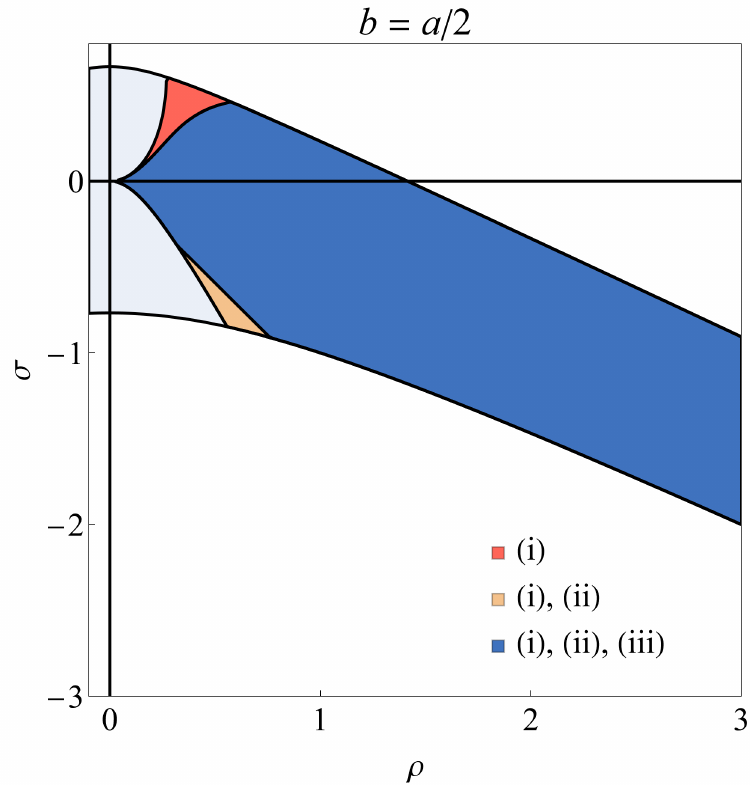}
\end{subfigure}
\hfill
\begin{subfigure}{0.48\textwidth}
    \centering
    \includegraphics[width=\textwidth]{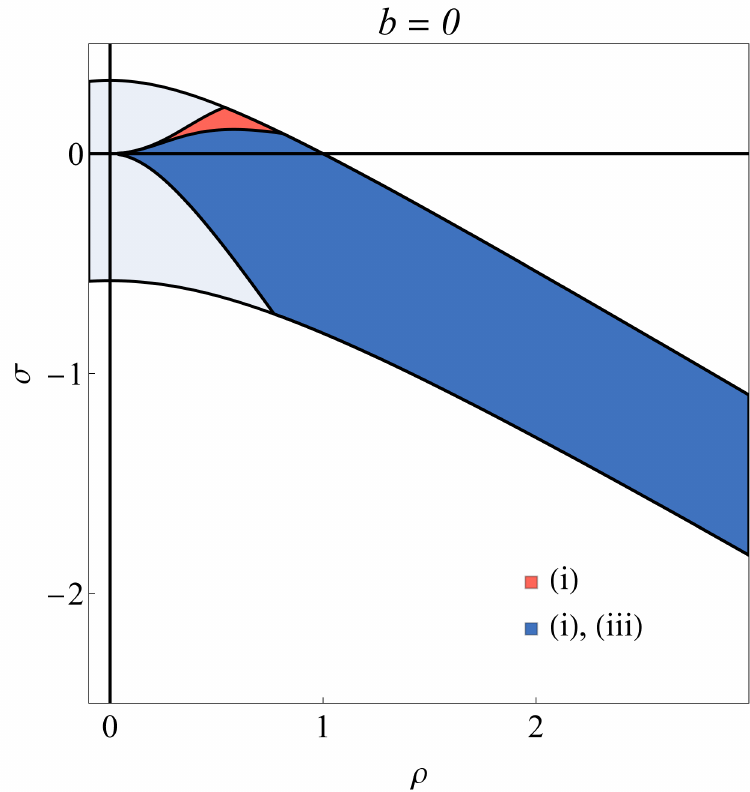}
\end{subfigure}
\caption{Classification of the $(\rho,\sigma)$ parameter space for $b=a/2$ and $b=0$. Each point is classified according to whether it (i) satisfies the convergence conditions of the SCI, (ii) satisfies the asymptotic KSW criterion, and (iii) satisfies both the near-horizon and asymptotic KSW criteria. The different regions correspond to the combinations of (i)–(iii) indicated in the legend.}
\label{fig:figure8}
\end{figure}

The asymptotic limit of the metric~\eqref{CCLPmetric} is obtained by first performing the change of coordinates:
\begin{align}
    \frac{\left(1-a^2\right) \sin ^2\vartheta }{z^2}=(r^2+a^2) \sin ^2\theta\, ,\qquad\frac{\left(1-b^2\right) \cos ^2\vartheta }{z^2}=(r^2+b^2) \cos ^2\theta\,, 
\end{align}
and then taking the limit $r \to \infty$. The asymptotic metric reads:
\begin{align}
    \diff s^2={}&\frac{\diff z^2}{z^2}+\frac{\beta ^2\, \diff \tau^2+\cos ^2\vartheta\, \left(\diff \psi -i \beta  \Omega_2 \,\diff \tau\right)^2+\sin ^2 \vartheta\, \left(\diff \phi-i \beta  \Omega_1 \,\diff\tau \right)^2+\diff \vartheta^2}{z^2}\,.
    \label{asymptotic-metric}
\end{align}
Only the $(\tau,\phi,\psi)$ part of the metric is complex-valued. We then apply the criterion to the reduced metric
\begin{align}
g_{(3)}=  \frac{1}{z^2} \left(
\begin{array}{ccc}
 -\beta ^2 \left(\Omega_1^2 \,\sin ^2\vartheta +\Omega_2^2\, \cos ^2\vartheta -1\right) & -i \beta \, \Omega_1 \sin ^2\vartheta & -i \beta \, \Omega_2 \cos ^2\vartheta \\
 -i \beta \, \Omega_1 \sin ^2\vartheta &\sin ^2\vartheta & 0 \\
 -i \beta\,  \Omega_2 \cos ^2\vartheta & 0 & \cos ^2\vartheta \\
\end{array}
\right).
\end{align}
The $p=0$ criterion as usual fixes $\re\,\beta >0$.
The $p=1$ condition, in turn, requires the matrix $A^{(3)}=\re \left[(\det g_{(3)})^{-1/2}g_{(3)}\right]$
to be positive-definite.\footnote{For complex symmetric matrices, $\re(M)>0$ if and only if $\re(M^{-1})>0$, so this is equivalent to the original condition $\re\left(\sqrt{\det g_{(3)}}\,g_{(3)}^{-1}\right)>0$.} This is equivalent to
\begin{align}\label{p=1_2}
&A^{(3)}_{22}>0\,,\qquad A^{(3)}_{33}>0\,, \qquad-\frac{\left(A^{(3)}_{12}\right)^2}{A^{(3)}_{22}}-\frac{\left(A^{(3)}_{13}\right)^2}{A^{(3)}_{33}}+A^{(3)}_{11}>0  \,.
\end{align}
The third inequality in~\eqref{p=1_2} is equivalent to
\begin{equation}
    \frac{z}{\cos \vartheta \sin \vartheta} \left(\frac{\sin ^2\vartheta\, \re(\beta -\beta \, \Omega_1)\, \re(\beta+\beta\,  \Omega_1 )+\cos ^2\vartheta \, \re(\beta -\beta \, \Omega_2) \,\re(\beta+\beta \,\Omega_2 )}{\re \,\beta}\right)>0\,.
\end{equation}
New constraints are imposed by~\eqref{p=1_2}, which is satisfied for all values of $\vartheta$ iff 
\begin{align}
\nonumber    &\re [\beta (1-\Omega_1)]\, \re [\beta(1+\Omega_1)]>0 \,, \qquad\re [\beta (1-\Omega_2)]\, \re [\beta(1+\Omega_2)]>0\,.
\end{align}
For $b=\pm a$ these are always satisfied for horizons within the index-convergence domain. In the other two cases, we find that they exclude a larger region of parameter space (red regions in figure~\ref{fig:figure8}).

As in the four-dimensional case, the additional asymptotic KSW
constraints coincide with~\eqref{eq:nonBPSdescendantconvergence},
which are required for convergence of the field theory partition function. In this case, these constraints are stronger than the corresponding index-convergence conditions when evaluated on a solution with $a\ne \pm b$. This is in contrast to what we have found in four-dimensions, and in the
family studied in \cite{BenettiGenolini:2026raa,Krishna:2026rma}. In that case the asymptotic KSW and index-convergence conditions turn out to be equivalent upon using the precise relations between the chemical potentials and the black hole parameters. This mismatch can be interpreted as a concrete indication that the KSW criterion might need some refinement in order to agree with the index convergence for generic complexifications.

\paragraph{Near-horizon KSW analysis.} In terms of the near-horizon coordinate $r-r_+=\kappa R^2$, the metric takes the form
\begin{align}
\nonumber    \diff s^2\,\sim\, &\frac{4 \Sigma(r_+)^2 \,\kappa}{\Delta_r'(r_+)}\,(\diff R^2+4 \pi^2R^2 \diff \tau^2)+\frac{\Sigma(r_+)^2}{\Delta_{\theta}} \diff \theta^2\\
    &+g_{\phi \phi} (r_+)\,\diff \phi^2+g_{\psi \psi}(r_+)\,\diff \psi^2+2g_{\phi\psi}(r_+)\, \diff \phi \, \diff \psi\,.
\end{align}
We consider the weaker version of the criterion obtained by making the overall coefficient of the $(R,\tau)$ block real and positive for each fixed value of $\theta$. At each fixed angle $\theta$, the KSW criterion therefore reduces to the $p=0$ and $p=1$ conditions on the transverse three-dimensional metric. The resulting conditions are then interpreted as being necessary for the full criterion to be satisfied. We evaluate these conditions numerically at representative values of the polar coordinate, $\theta\simeq0,\pi/4,\pi/2$. Strictly speaking, $\theta=0$ and $\theta=\pi/2$ are coordinate-degenerate points of the $S^3$ parametrization, and therefore the numerical analysis is performed at points arbitrarily close to them rather than exactly at the endpoints.
The $p=0$ condition is always satisfied by the outer horizons. For generic $a$ and $b$, after imposing $\im \, m_+=0$, it can be written as:
\begin{equation}
    \re \left( \sqrt{\det g_{S^3}}\right)=\pm\frac{\rho  \sin 2 \theta  \left(a^2 (b+1)+a b (b-2 \sigma +1)+b^2+\rho ^2-3 \sigma ^2\right)}{2 \left(a^2-1\right) \left(b^2-1\right)}>0\,.
\end{equation}
For $\rho\neq0$, the sign of the square root can always be chosen so that this inequality is satisfied. More precisely, the positive branch is selected for $\rho>0$ ($\rho<0$) on the first (second) horizon branch, while the negative branch applies for $\rho<0$ ($\rho>0$), respectively.

The non-trivial restriction arises from the $p=1$ condition, which requires the matrix $A^{nh}=\re \left[\left(\det g_{S^3}\right)^{-1/2}g_{S^3}\right]$
to be positive definite. This implies
\begin{align}
A^{nh}_{\theta \theta}>0\,,\qquad A^{nh}_{\phi \phi}>0\,,\qquad A^{nh}_{\psi \psi}-\frac{\left(A^{nh}_{\psi \phi}\right)^2}{A^{nh}_{\phi \phi}}-\frac{\left(A^{nh}_{\psi \theta}\right)^2}{A^{nh}_{\theta \theta}}>0\,.
\end{align}
 Among the sampled angles, the strongest restrictions are obtained at $\theta\simeq\pi/2$. The regions satisfying both the near-horizon and asymptotic KSW conditions at this angle are shown in figure~\ref{fig:figure8} in blue. 

On the second branch, the $p=0$ and $p=1$ KSW conditions are never simultaneously satisfied. Hence, the near-horizon KSW criterion excludes it completely. The constraints imposed on the first branch are instead weaker, with a small additional set of previously allowed solutions that gets removed only for $b=a/2$ (yellow region in figure~\ref{fig:figure8}).

\raggedbottom
 \paragraph{Bulk KSW analysis.} We conclude with the analysis of the KSW criterion throughout the bulk geometry. 
As in the four-dimensional case, we find that the bulk analysis provides additional constraints in all four cases we consider.

\begin{figure}[H]
\begin{subfigure}{0.48\textwidth}
    \centering
    \includegraphics[width=\textwidth]{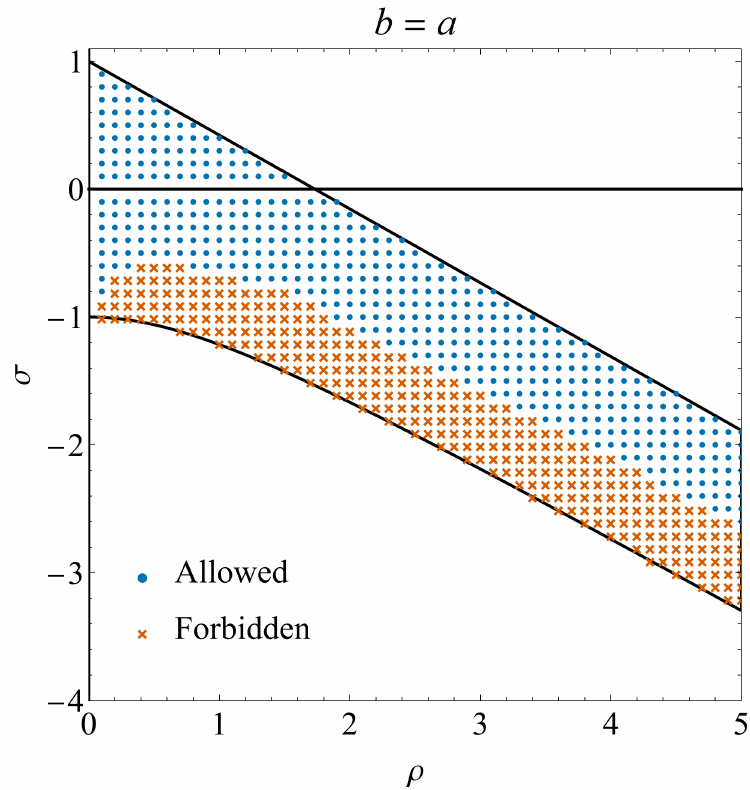}
\end{subfigure}
\hfill
\begin{subfigure}{0.48\textwidth}
    \centering
    \includegraphics[width=\textwidth]{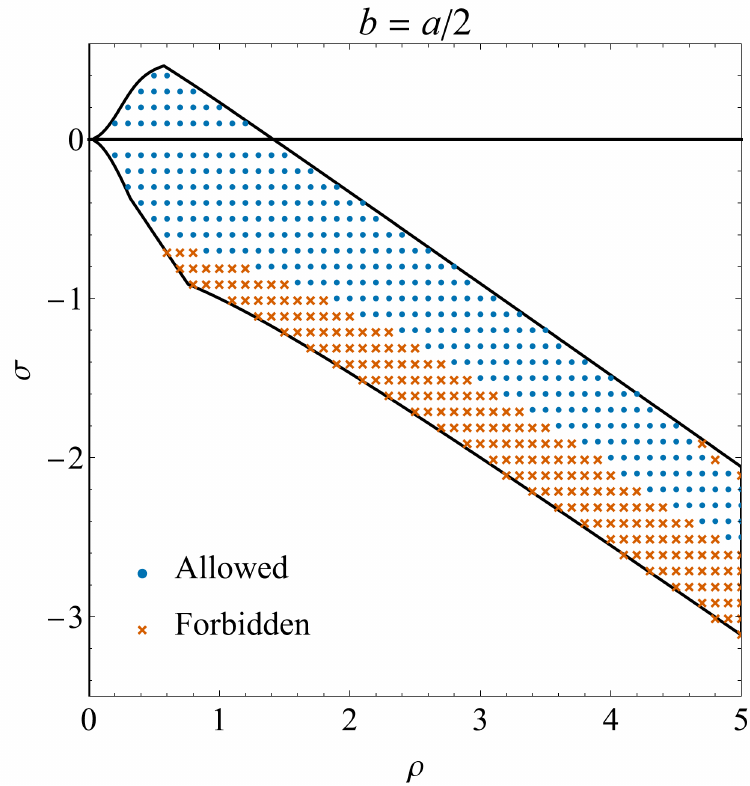}
\end{subfigure}

\begin{subfigure}{0.48\textwidth}
    \centering
    \includegraphics[width=\textwidth]{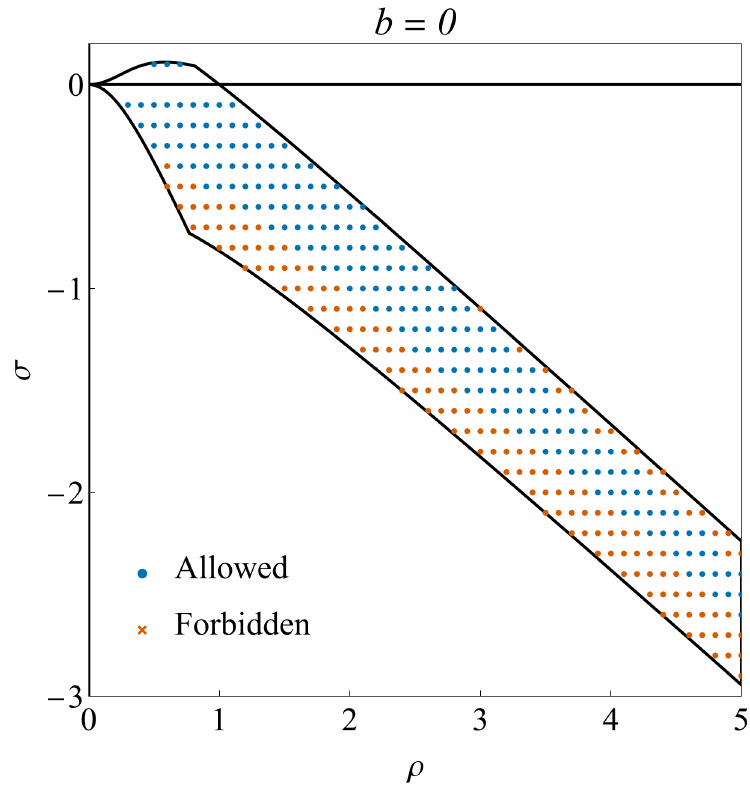}
\end{subfigure}
\hfill
\begin{subfigure}{0.48\textwidth}
    \centering
    \includegraphics[width=\textwidth]{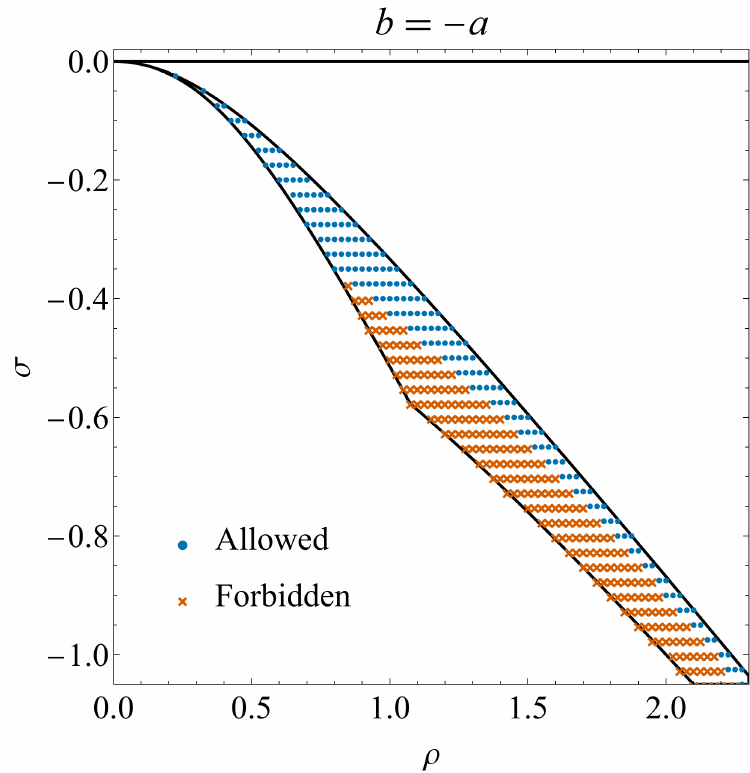}
\end{subfigure}
\caption{Numerical classification of the first branch of outer horizons in the $(\rho,\sigma)$-plane according to the bulk KSW criterion, for different choices of the angular-momentum parameter $b$. 
The solid black curve delimits the portion of the horizon domain satisfying the near-horizon and asymptotic KSW admissibility constraints.}
\label{fig:figure9}
\end{figure}

We proceed as in section~\ref{sec:3.3}, applying the third formulation of the KSW criterion to the reduced four-dimensional metric obtained by removing the $g_{rr}$ component. We determine whether there exist contours connecting the horizon to the asymptotic region along which the reduced metric can be KSW-allowable. Again, this provides a necessary condition for a given horizon to be allowed.
The most stringent condition is attained for $\theta \simeq \frac{\pi}{2}$, and the results of the numerical analysis for this value of $\theta$ are shown in figure~\ref{fig:figure9}. Also in this case, the bulk analysis removes a large portion of previously allowed horizons that lie close to the boundaries of the horizon domain. In this case, such boundaries correspond to (singular) solutions with $a=\pm1$.

\section{Summary and discussion}\label{sec:discussion}
In this work, we introduced a new complexification of supersymmetric non-extremal black holes that preserves real conserved charges while allowing the radial coordinate and the horizon position to become complex. This prescription explores a section of the supersymmetric solution space distinct from other prescriptions that appeared in the literature \cite{Cabo-Bizet:2018ehj,Cassani:2019mms,Bobev:2019zmz}.
 
For this family of solutions, the field-theory and gravitational conditions do not generally coincide. In five dimensions, the asymptotic KSW criterion is more restrictive than index convergence, excluding saddles whose boundary chemical potentials lie within the convergence domain of the index. We interpret this mismatch as due to the fact that the KSW criterion, in the formulation considered in this paper, is not adapted for a supersymmetric setup. On the one hand, it only takes into account the metric without considering other background fields, such as the gauge potential, which are necessary in order to realize supersymmetry. On the other hand, it constrains a generic probe quantum field theory defined on the fixed background, without incorporating the supersymmetric structure responsible for the cancellations entering the index.
It would therefore be interesting to investigate whether the criterion admits a refinement that captures the restriction to the BPS sector.

In both four and five dimensions, the bulk KSW analysis imposes further restrictions, excluding additional saddles that satisfy the asymptotic criterion. While an asymptotic violation may be associated with an ill-defined grand-canonical ensemble, a bulk violation may instead indicate that the chosen filling is inadmissible despite satisfying the boundary conditions. This raises the possibility that another bulk saddle with the same boundary conditions provides an admissible solution. A transition from bulk allowability to non-allowability could then be associated with a change in the relevant saddle, suggesting a possible connection between the bulk KSW criterion and phase transitions. 
Establishing such a connection would require identifying the competing saddles and comparing their contributions to the gravitational path integral.  

We conclude by discussing two alternative prescriptions for selecting complex saddles, whose comparison with the KSW criterion may help determine whether the restrictions obtained from the gravitational analysis are reproduced by independent methods.

One possibility is to compare the KSW condition with the non-perturbative criterion proposed in~\cite{Aharony:2021zkr,Suh:2026ikt}. Uplifting  the AdS$_4$ black hole to M-theory, one may consider an M5-brane wrapping an $S^5\subset S^7$ factor and the $S^1$ Euclidean time circle in the AdS$_4$ geometry. The corresponding action was evaluated explicitly in~\cite{Suh:2026ikt}. Such brane configurations lead to a divergence of the gravitational path integral unless
\begin{align}\label{braneaction}
 \im (S_{M5})=\im \left(\mp 64\pi N \frac{\varphi}{\omega}\right)>0\,,
\end{align}
where the sign depends on the branch chosen in~\eqref{susy-chemical-constraint}. For the upper-sign branch, the condition~\eqref{braneaction} reduces simply to $\re \, \omega<0$. The probe-brane criterion reproduces part of the field-theory convergence conditions and is less restrictive than the KSW criterion. Analogous considerations hold in the five-dimensional case.

Another possible approach is a Picard-Lefschetz analysis within a minisuperspace approximation, recently applied to black holes in \cite{Mahajan:2025bzo,Singhi:2025rfy,Kolanowski:2026ylt}. Using this framework, the gravitational path integral is reduced to an integral over a finite set of parameters, such as \(a\) and \(r_+\), allowing one to determine which saddles contribute to a specified integration contour. Realizing such an analysis for the complexification considered here would, however, require additional care. As discussed in section~\ref{sec:charged-rotating-ads4}, this complexification keeps the chemical potentials at their BPS values while varying the inverse temperature. The independence of the supersymmetric on-shell action of $\beta$ at fixed reduced chemical potentials suggests that an analysis restricted to the supersymmetric locus may not capture all relevant features of this deformation. A possible approach would therefore be to relax the supersymmetry constraint, restoring the nontrivial dependence of the action on $\beta$, and investigate whether these saddles contribute to the gravitational path integral.
\section*{Acknowledgments}
We would like to thank Federico Arrighi, Oliver Janssen, Alejandro Ruipérez, Enrico Turetta and especially Davide Cassani for discussions. We also thank Federico Arrighi and Davide Cassani for comments on the draft. We thank the Galileo Galilei Institute for Theoretical Physics for hospitality during the completion of this work.

\appendix
\section{Extremization principle with real charges}
\label{app:RealChargExtrPrinc}
In this appendix we prove that reality of the conserved charges restricts the chemical potentials of general supersymmetric and non-extremal black holes to the BPS locus. The argument relies on the extremization procedure that yields the black hole entropy from the Euclidean on-shell action $I$, while keeping real charges.

\paragraph{Four-dimensional black hole.} Upon imposing supersymmetry, the Euclidean on-shell action of the four-dimensional black hole discussed in section~\ref{sec:charged-rotating-ads4} reads \cite{Cassani:2019mms,Bobev:2019zmz}
\begin{equation}
    I = \pm \frac{i}{2} \frac{\varphi^2}{\omega} \,, \qquad \omega-2\varphi=\pm2\pi i\,,
\end{equation}
where in the two equations the signs are related. Importantly, we assume that the chemical potentials $\omega,\varphi$ are tuned in such a way to describe a solution with real charges. We can explicitly see what this condition implies by expressing the chemical potentials as functions of the conserved charges $J,Q$. This can be achieved by extremising the following function with respect to $\omega,\varphi$
\begin{equation}
    S = \text{ext}_{\omega,\varphi}\bigl[-I -\omega\,J-\varphi\,Q +\Lambda(\omega-2\varphi\mp2\pi i)\bigr]\,,
\end{equation}
where the Lagrange multiplier $\Lambda$ enforces the constraint on $\omega$ and $\varphi$. From the extremization equations one finds
\begin{equation}
    \omega = -\frac{2\pi}{4\Lambda+2Q\pm i}\,, \qquad S =\mp 2\pi i\,\Lambda\,,
    \label{eq:4dExtrEqsSols}
\end{equation}
where $\varphi$ is fixed by the linear constraint and $\Lambda$ is fixed by the equation
\begin{equation}
    (Q+2\Lambda)^2 \pm2i(\Lambda-J) = 0\,,
    \label{eq:4dEqForLambda}
\end{equation}
which generally leads to complex solutions for $\Lambda$, corresponding to complex values for the entropy $S$. By writing $\Lambda = \re\,\Lambda\pm i\,\im\,\Lambda$, with sign choice again related to the one made in the action, and under the assumption that the charges $Q,J$ are real, eq.~\eqref{eq:4dEqForLambda} reduces to a set of two equations for $\re\, \Lambda$ and $\im\,\Lambda$
\begin{equation}
    \begin{cases}
        (2J+Q)^2 = X^2(1+4X^2)\,, \\[2mm]
        \im\, \Lambda = \frac{2J+Q-X}{4 X}\,,
        \end{cases}
        \label{eq:4dEqForLambda2}
\end{equation}
where we introduced $X = 2\, \re \,\Lambda+Q$ for convenience. The extremal BPS solutions are obtained by further imposing that the entropy is real, from~\eqref{eq:4dExtrEqsSols} this implies $\re \,\Lambda^\star=0$ and hence $X^\star=Q^\star$. As the charges are taken real, no further reality conditions have to be imposed and~\eqref{eq:4dEqForLambda2} reduces to the familiar non-linear constraint \cite{Cassani:2019mms,Bobev:2019zmz} that imposes extremality of the solution.

Coming back to the general non-extremal solutions, by making use of~\eqref{eq:4dEqForLambda2} one can show that $\omega$ can be rewritten solely as a function of $X$
\begin{equation}
    \omega = -\frac{4\pi X}{1+8X^2} \pm 2\pi i\frac{\sqrt{1+4X^2}}{1+8X^2}\,.
\end{equation}
This implies that $\omega$ does not depend on the independent real charges $Q,J$ separately but only through their combination associated with $X$. As a consequence, the values acquired by $\omega$ lie on a one-dimensional curve in the complex $\omega$ plane as $J$ and $Q$ independently vary. This curve corresponds to the BPS locus, i.e. the one-dimensional locus defined by the values $\omega^\star$ of the chemical potential attained by BPS solutions. This can be seen by considering that $\omega^\star$ is given by
\begin{equation}
    \omega^\star = -\frac{4\pi Q^\star}{1+8Q^{\star\,2}} \pm 2\pi i\frac{\sqrt{1+4Q^{\star\,2}}}{1+8Q^{\star\,2}}\,.
\end{equation}
Note that this does not imply that the supersymmetric solutions are BPS solutions. The relation $\omega\bigl(X(Q,J)=Q^\star\bigr) = \omega^\star(Q^\star)$ is valid for every supersymmetric configuration for which $X(Q,J)=Q^\star$. The BPS condition would  require $X(Q^\star,J^\star)=Q^\star$ which instead imposes the extremal non-linear constraint on the charges.

\paragraph{Five-dimensional black hole.} The proof for the five-dimensional case is complicated by the presence of an additional charge. We explicitly prove our result for the $J_1=J_2 =J$ case.

The Euclidean on-shell action for the five-dimensional supersymmetric black hole discussed in section~\ref{sec:5d} reads \cite{Cabo-Bizet:2018ehj} (fixing $J_1=J_2=J$)
\begin{equation}
    I = \frac{2\pi}{27}\frac{\varphi^3}{\omega^2}\,, \qquad \omega-\varphi=\mp \pi i\,,
\end{equation}
the Bekenstein-Hawking entropy is obtained as a function of the charges via:
\begin{equation}
    S = \text{ext}_{\omega,\varphi}\bigl[-I -2\omega J-\varphi Q +2\Lambda(\omega-\varphi\pm\pi i)\bigr]\,.
\end{equation}
From the corresponding extremization equations one can extract $\omega$ as a function of $Q,J,\Lambda$
\begin{equation}
    \omega= \mp i\pi\frac{Q+2\Lambda}{3J+Q-\Lambda}\,,
\end{equation}
while the equation for the Lagrange multiplier $\Lambda = \re \,\Lambda\pm i \,\im\, \Lambda$ can be reduced to the following two equations (introducing $2\re\,\Lambda =X-Q$)
\begin{equation}
    \begin{cases}
        4\,(\im\,\Lambda)^2 =3X^2+\pi(X-Q-2 J)\,, \\[2mm]
        16\,(\im\,\Lambda)^4-4\,(\im\,\Lambda)^2\bigl(6X^2+6\pi X+\pi^2\bigr)+X^3\bigl(9X+2\pi\bigr)=0\,.
    \end{cases}
    \label{eq:5dEqsLambda}
\end{equation}
Using the first equation allows us to write $\omega$ solely as a function of $\im \,\Lambda$ and $X$
\begin{equation}
    \omega = \pm i\frac{2\pi^2\bigl(X\pm2 i\,\im\,\Lambda\bigr)}{12\,(\im\,\Lambda)^2\pm2 \pi i\,\im\,\Lambda-X(9X+2\pi)}\,,
\end{equation}
then, using the second equation in~\eqref{eq:5dEqsLambda} allows to rewrite $\omega$ as a function of only $X$. As before, $\omega$ only explicitly depends on a given combination of the independent real charges $J,Q$. Moreover, the BPS limit is again reached for $\re\,\Lambda=0$ which implies $X^\star=Q^\star$. It follows that also in this case the chemical potentials take values only on the BPS locus.
In the $J_1 \ne J_2$ case, one should similarly find that the chemical potentials of the supersymmetric but non-extremal solutions can be written in terms of only two variables $X,Y$ of the three independent real charges $J_1,J_2,Q$.

\bibliographystyle{JHEP}
\bibliography{KSW.bib}
\end{document}